\documentclass[acmsmall, screen]{acmart}

\PassOptionsToPackage{dvipsnames}{xcolor}
\PassOptionsToPackage{colorlinks=true,linkcolor=RubineRed,breaklinks=True,citecolor=RubineRed}{xcolor}

\usepackage{enumitem}% http://ctan.org/pkg/enumitem
\setlist[itemize]{noitemsep, topsep=0pt}

\usepackage{colortbl}
\usepackage[most]{tcolorbox}
\usepackage{adjustbox}
\usepackage[linesnumbered,algo2e,ruled]{algorithm2e}
\usepackage{multirow}
\usepackage{booktabs,subcaption,dcolumn}
\usepackage{tikz}
\usepackage{enumitem}
\usepackage{tabularx}

\usepackage{pifont}% http://ctan.org/pkg/pifont
\usepackage{svg}
\usepackage{soul}
\usepackage{amsmath}
\usepackage{dirtytalk}
\usepackage{graphicx}
\usepackage{setspace}
\usepackage[font=small]{caption}
\usepackage{booktabs} %toprule / bottomrule 
\usepackage{svg}
\usepackage{amsfonts}
\usepackage{xurl} 

\usepackage{diagbox}%y 18 mar
\usepackage{indentfirst}
\usepackage{xspace}
\input{commands}

\AtBeginDocument{%
  }

\setcopyright{acmlicensed}
\acmJournal{TWEB}
\acmYear{2025} \acmVolume{1} \acmNumber{1} \acmArticle{1} \acmMonth{1}\acmDOI{10.1145/3770852}

\begin{document}

\title{It's not Easy: Applying Supervised Machine Learning to Detect Malicious Extensions in the Chrome Web Store}

\author{Ben Rosenzweig}
\email{s8berose@stud.uni-saarland.de}
\orcid{0009-0008-6556-3230}
\affiliation{%
  \institution{Saarland University}
  \city{Saarbruecken}
  \country{Germany}
}

\author{Valentino Dalla Valle}
\email{valentino.dalla-valle@cispa.de}
\orcid{0009-0008-3557-3120}
\affiliation{%
  \institution{CISPA Helmholtz Center for Information Security}
  \city{Saarbruecken}
  \country{Germany}
}

\author{Giovanni Apruzzese}
\orcid{0000-0002-6890-9611}
\email{giovanni.apruzzese@uni.li}
\affiliation{%
  \institution{University of Liechtenstein}
  \city{Vaduz}
  \country{Liechtenstein}
}
\affiliation{%
  \institution{Reykjavik University}
  \city{Reykjavik}
  \country{Iceland}
}

\author{Aurore Fass}
\email{fass@cispa.de}
\orcid{0000-0001-6611-4447}
\affiliation{%
  \institution{CISPA Helmholtz Center for Information Security}
  \city{Saarbruecken}
  \country{Germany}
}

\begin{abstract}
Google Chrome is the most popular Web browser.
Users can customize it with \textit{extensions} that enhance their browsing experience. The most well-known {marketplace} of such extensions is the Chrome Web Store (CWS). Developers can upload their extensions on the CWS, but {such extensions} are made available to users only after a vetting process carried out by Google itself. Unfortunately, some \textit{malicious} extensions bypass such checks, putting the security and privacy of downstream {browser extension} users at risk. 

In this paper, we carry out a comprehensive real-world security analysis of malicious extensions in the CWS.
Specifically, we scrutinize the extent to which automated mechanisms reliant on supervised machine learning (ML) can be used to detect malicious extensions on the CWS.
To this end, we first collect 7,140 malicious extensions published in 2017--2023 and which have been flagged as {malicious} by Google.
We combine this dataset with {63,598} benign extensions published or updated on the CWS before 2023, and we develop three supervised-ML-based classifiers---leveraging both original features as well as techniques inspired by prior work.
We show that, {in a ``lab setting'', our} classifiers work well (e.g., 98\% accuracy).
Then, we collect a new, and more recent, set of 35,462 extensions from the CWS, published or last updated in 2023, with unknown ground truth.
We were eventually able to identify 68 malicious extensions that bypassed the vetting process of the CWS. However, our classifiers also reported over 1k likely malicious extensions which may overestimate their true number.
Based on this finding (further supported with other experiments and realistic analyses), we elucidate, for the first time, a strong \textit{concept drift} effect on browser extensions.
We also provide factual evidence that commercial detectors (e.g., VirusTotal) work poorly to detect known malicious extensions.
Altogether, our results highlight the fact that detecting malicious browser extensions is a fundamentally hard problem which has not (yet) received an adequate degree of attention. This requires additional work both by the research community and by Google itself---potentially by revising their approaches. In the meantime, we informed Google of our discoveries, and we release our artifacts.
\end{abstract}




% \received{20 February 2025}
% \received[revised]{day month year}
% \received[accepted]{day month year}

\maketitle

\section{Introduction}
\label{sec:introduction}

\noindent
Web browsers have become indispensable tools: besides allowing users to visit and interact with various websites, browsers now integrate a rich ecosystem of \textit{browser extensions} that enhance their functionalities. Such add-ons can substantially improve users' online experience---e.g., by removing ads, facilitating password management, or automatically spell-checking users' input~\cite{Hsu2024}. In this context, the Chrome Web Store (CWS) stands out as the most popular ``marketplace'' of extensions supported by chromium-based Web browsers~\cite{cws}: as of {September} 2025, several hundreds of extensions in the CWS are installed by millions of devices worldwide~\cite{topextensions}.  

Unfortunately, despite most extensions providing plenty of benefits to their users, some extensions may conceal harmful content that can lead to security (e.g., unauthorized operations~\cite{kharraz2019outguard}) or privacy (e.g., data leakage~\cite{Chen2018}) violations. Such \textit{malicious extensions} are a cause of concern to Web users, who may inadvertently compromise their devices by installing a malicious extension from a legitimate source---i.e., the CWS. 
Indeed, the CWS adopts an (apparently) strict policy for the content uploaded by third-party developers: before becoming available to end-users, any new uploaded extension must first pass Google's vetting process~\cite{CWSreviewprocess, review-status, staying-safe}. Despite such precautionary measures, a recent work~\cite{Hsu2024} found that the CWS is rich of ``security-noteworthy extensions'', which bypass the checks enforced by Google, and hence represent a risk for any users installing them onto their Web browser.

Abundant prior work attempted to mitigate the spread of malicious browser extensions~\cite{Wang2018, Aggarwal2018, Pantelaios2020}, either by assessing their real-world prevalence (e.g.,~\cite{Jagpal2015, Hsu2024}) or by studying the risks of certain types of extensions, such as those exfiltrating user data~\cite{Chen2018, Xie2024, Starov2017-2} (potentially by also proposing some countermeasures~\cite{Weissbacher2017,Pantelaios2020}).
Yet, we found no recent work that carried out a comprehensive, reproducible, and large-scale security analysis of classifiers reliant on supervised machine learning (ML) to automatically detect malicious browser extensions on the CWS. 
For instance, some works analyze only hundreds of extensions~\cite{Carlini2012}, whereas others do not focus on the CWS~\cite{Wang2012,zhao2021privacy}, are outdated (e.g., 2012--2015~\cite{Jagpal2015, Kapravelos2014, Wang2012, Xing2015, kurt2015}), or propose methods reliant on unsupervised ML techniques---and therefore the detection process requires defining rules/heuristics or manual inspections (e.g.,~\cite{Pantelaios2020}). We systematically discuss related work in Section~\ref{sec:related}.
Intriguingly, however, there is abundant literature that performs holistic analyses focused on the detection of other (and orthogonal) security threats---such as generic Android/Windows malware~\cite{cai2020assessing,pendlebury2019tesseract,galloro2022systematical} or Phishing Email/Webpages~\cite{oest2020phishtime, cui2017tracking, lain2022phishing}. This shows that the detection of malicious browser extensions by means of automated methods is still an open problem requiring further scrutiny.

In this paper, we tackle this challenge and carry out a security assessment of the extensions published on the CWS. We begin with a broad research question: can we build a system that automatically learns (via supervised ML) to detect malicious extensions uploaded on the CWS?
To this end, we first collect a \textit{large} corpus of over \textit{100k} extensions: 99k are from the CWS (and we are unsure of their true nature), whereas 7k are \textit{verified malicious} extensions taken from Chrome-Stats~\cite{ChromeStats}; these procedures are outlined in \Cref{sec:data}.
Then, we use these extensions to develop three automated detectors---based both on original features and prior work~\cite{JaSt}, used to train supervised ML-based classifiers. We show that these detectors achieve remarkable performance, with 98\% accuracy and less than 2\% false positives, by carrying out rigorous experiments and following best practices~\cite{arp2022and} (including cross validation and verification of the feature importance); this evaluation is discussed in \Cref{sec:labeled-m-ext-detect}.
Based on these encouraging results, we test our detectors on extensions in the ``real world,'' i.e., on extensions recently (2023) published on the CWS, for which we have no verified ground truth. These experiments are discussed in \Cref{sec:open} and reveal a stark disconnection with our initial findings. First, our detectors flagged over 1k extensions (out of $\sim$35k tested) as malicious. Then, we manually analyzed a subset {($\approx$20\%)} of them, and we find that \tk{only 68} of these extensions can be claimed to be {truly} malicious---denoting a large number of false positives. 
This result suggests that the ecosystem of browser extensions may be affected by the \textit{concept drift} problem. We validate this hypothesis in \Cref{sec:concept}, wherein we simulate a realistic (and fair) longitudinal analysis, measuring how the performance of our detectors would have changed over four years. Our results confirm the existence of concept drift, elucidating that {automatic detection of} malicious extensions is fundamentally hard. We even tested commercial products, such as VirusTotal, and found they work poorly: they cannot detect known malicious extensions taken down by Google.

\noindent
\textsc{\textbf{Contributions}.} 
This paper provides technical methods and novel scientific findings. Specifically:
\begin{itemize}[leftmargin=*]
    \item we collect $\approx$107k browser extensions and use them to develop, and then evaluate, three detectors (including original ones) of malicious browser extensions;
    \item we show that---in our ``laboratory'' setup---our detectors perform well (detection accuracy=98\%, $\approx$1s to analyze an extension end-to-end) which would justify their deployment in the real world;

    \item by using our detectors in an open-world test on 35k extensions, we provide (for the first time) factual evidence that the CWS ecosystem is affected by concept drift, making automated detection hard in the long term. 
    We analyze the effects of concept drift and explore some mitigations;
    
    \item nonetheless, by manually analyzing the extensions flagged by our detectors, we found 68 malicious extensions on the CWS, which affected $>$13M users. We disclosed our findings to Google.
\end{itemize}
\noindent
To pave the way for future research, we discuss the lessons learned from our security analysis in Section~\ref{sec:conclusions}. We publicly release our artifacts~\cite{repository} for reproducibility.
\section{Background: Browser Extensions}
\label{sec:background}

\noindent
Browser extensions are third-party programs that users can install to improve their browsing experience by, e.g., removing advertisements from Web pages or for password management.
To set up the stage for our contributions, we first describe the generic architecture of a browser extension and then outline its main components. Given the fact that we focus on the CWS, the following content pertains to extensions for chromium-based Web browsers.

\subsection{Architecture of a Browser Extension}
\label{subsec:architecture}
\noindent
Browser extensions are \texttt{crx} files, i.e., compressed archives including mainly HTML, JavaScript, and CSS files.
Every extension must have a \texttt{manifest.json} file~\cite{cat}. The \texttt{manifest} contains basic configuration and details about an extension, such as its main components and permissions.
The current version (as of September 2025) is \textit{Manifest V3} which aims to improve the security of extensions by, e.g., preventing them from downloading external resources~\cite{ManifestV3}. Although the CWS stopped accepting V2 extensions in January 2022, only 76.8\% of the extensions currently in the CWS have migrated to V3~\cite{MigrationV3}.

{\tt \tiny
\begin{lstlisting}[language=json, caption=Extract of a \texttt{{\footnotesize manifest.json}} file from a browser extension, float, label=lst:manifest]
"manifest_version": 3,
"permissions": ["downloads", "history"],
"host_permissions": ["https://example.com/*"],
"background": {
    "service_worker": "service_worker.js",
},
"content_scripts": [
    {
        "matches": ["<all_urls>"],
        "js": ["script.js"]
    }
],
...
\end{lstlisting}

}

To use most extension APIs (e.g., \texttt{cookies}, \texttt{downloads}, \texttt{history}), a developer must declare the permissions in the corresponding \texttt{manifest} field~\cite{Declare-permissions, Permissions-list}.
In addition, a developer can also specify host permissions (i.e., URLs or URL patterns) to allow extensions to interact only with specific Web pages (useful to, e.g., monitor network requests or fetch data). 
An example of a manifest file is shown in \Cref{lst:manifest}. Here, the extension has the permissions ``downloads'' and ``history'' to download arbitrary files and read or modify users' browsing history.
Through the host permissions, it is able to send HTTP requests to ``https://example.com/*''.

\subsection{Main Components of a Browser Extension}
\label{subsec:components}

\begin{figure}[t]
    \centering
    \includegraphics[width=.5\columnwidth]{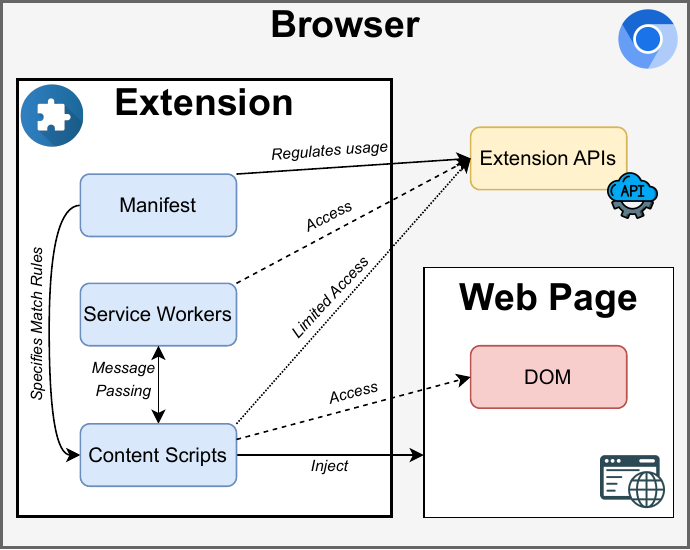}
    \vspace{-1mm}
    \caption{Typical architecture of a chromium-based browser extension}
    \label{fig:architecture}
\end{figure}

\noindent
\Cref{fig:architecture} displays the architecture of browser extensions and their main components relevant to this work.
The core logic of an extension is implemented through a \textit{service worker} (or \textit{background scripts} in Manifest V2). This JavaScript file runs in the background of the browser (independently of the lifetime of a window\,/\,tab), handles events, and has access to the extension APIs specified in the \texttt{manifest}~\cite{migrateToSW}.
An extension can inject \textit{content scripts} in a Web page. For example, in \Cref{lst:manifest}, a content script ``script.js'' is defined and injected into any Web page (matching pattern ``\textless~all\_urls~\textgreater'') visited by the user.
Similarly to scripts already loaded by Web pages, content scripts can use the standard DOM APIs to read and modify Web pages. Contrary to the service worker, content scripts have restricted access to the extension APIs (limited to \texttt{storage}, \texttt{i18n}, and various \texttt{runtime} APIs~\cite{Content_Scripts}). 

\subsection{Problem Scope, Motivation, and Threat Model}
\label{subsec:scope}

\noindent
The starting point of our research is developing detectors (i.e., binary classifiers) for malicious browser extensions that leverage supervised machine learning (ML) techniques. This goal is motivated by the intrinsic benefit of supervised ML methods to ``automatically'' learn hidden patterns among data for classification tasks---thereby avoiding the burden of manually defining ``rules'' or ``heuristics'' that can discriminate whether any given extension is benign or malicious~\cite{apruzzese2022sok}. Such learning is made possible during a training phase necessitating labeled data, used to define the decision boundaries of the ML model. Even though obtaining such labeled data is typically not trivial in cybersecurity~\cite{apruzzese2022sok}, the browser-extension context facilitates such a data-collection process because it is possible to use extensions \textit{taken down from the CWS for being ``malicious''} as samples with verified ground truth (we will better explain our rationale in Section~\ref{subsec:datasets}). 

We observe that our focus on supervised ML implicitly puts works on unsupervised- (e.g., clustering~\cite{Pantelaios2020}) or heuristic-based methods (e.g.,~\cite{Kapravelos2014}) outside our scope. Indeed, even though many prior works proposed methods to identify malicious browser extensions, there is no publicly available implementation of a detector that relies on supervised ML techniques specifically designed for malicious browser extensions. We extensively discuss related work in Section~\ref{sec:related}, wherein we also perform a systematic literature review to validate this claim. The closest work we could find---which we use as a baseline---is \jast{}, a detector of malicious JavaScript that relies on supervised ML, but which is not designed to work (and has never been tested) on browser extensions.

To devise our detectors, we assume the following threat model. First, the attacker wants to develop (and massively release) a browser extension that, if installed by (unaware) users in their Web browsers, would enable the attacker to carry out some security violation (e.g., exfiltrating user data). To maximize the reach of such a malicious extension, the attacker wants to upload it on popular (and, ideally, trusted) repositories---such as the CWS. We assume the attacker may leverage simple techniques, such as code obfuscation~\cite{moog2021}, to conceal the malicious functionality of the browser extension and/or delay the time required to flag the extension as being malicious. Hence, to detect such malicious extensions, we will leverage three complementary sources of information: JavaScript code, which is a crucial component of browser extensions and can embed some malicious functionality; as well as an extension's \texttt{manifest}, \texttt{service worker}, and \texttt{content scripts}, since prior work showed that such elements can conceal traces of malicious behavior~\cite{Pantelaios2020, Kapravelos2014, vasiliadis2023writ, karami2021awakening}. We do not seek to build a detector that is specifically designed to withstand targeted evasion attempts.

Finally, we emphasize that our focus is on ``malicious'' extensions---which are orthogonal to other classes of security-noteworthy extensions~\cite{Hsu2024}, such as vulnerable (e.g.,~\cite{fass2021doublex}) or fingerprintable (e.g.,~\cite{Agarwal2024}) extensions. An extended overview of related works is deferred to Section~\ref{sec:related}.
\section{Data Collection and Preparation}
\label{sec:data}

\noindent
As a starting point, we describe how we assembled the extensions that enabled our security analysis.
We first explain our data collection procedure. Then, we describe how we preprocessed our data. Finally, we present how we derive the two main datasets used in our investigation.

\subsection{Collecting Extensions}
\label{subsec:data-collection}

\noindent
We are not aware of any publicly available and (relatively) up-to-date datasets with benign and malicious extensions that are suitable for our purposes. Hence, for a meaningful assessment (and to comply with our threat model), we create our datasets by retrieving chromium-based browser extensions from two trusted sources: the Chrome Web Store~\cite{cws} and Chrome-Stats~\cite{ChromeStats}.

\subsubsection{Chrome Web Store}
The CWS is the most well-known marketplace for chromium-based extensions and is maintained by Google itself. We use the CWS sitemap~\cite{Sitemap} to collect the URLs of \textit{all} available extensions. From such URLs, we extract the unique 32-character long IDs of all the extensions, which enables us to directly download these extensions. Specifically, we perform these operations on Nov. 16, 2023. We identified 168,953 extension IDs, but 10,829 extensions were either not available for download or required payment. Hence, we downloaded the remaining 158,124 extensions; their last update was made between 2012 and 2023. 

\subsubsection{Chrome-Stats}
\label{subsubsec:chrome-stats}
Chrome-Stats~\cite{ChromeStats} is a tool that provides metadata (e.g., last update, user count, description, or user reviews) as well as the source code of extensions that are---or were---in the CWS.
Chrome-Stats engineers developed a crawler that has been automatically collecting extensions from the CWS and extracting their metadata from the CWS, once a day, since July 5, 2020.
We use Chrome-Stats, because it also stores data of extensions that have been removed from the CWS.
By using the ``Advanced Search Functionality''~\cite{AdvancedSearch}, we can look for extensions that have been removed from the CWS due to being malware (with the filter option ``obsoleteReason'' and the value ``malware''). Note that those extensions have  been flagged as malicious by Google engineers, and taken down from the CWS.
Our search (done on Nov. 16, 2023) yielded 10,814 malicious extensions, but 3,131 of these were not available for download.\footnote{{Chrome-Stats told us some developers asked for their extensions' source code to be removed from Chrome-Stats' database.}} Hence, we downloaded all the remaining 7,683 \textit{malicious} extensions. Finally, for every downloaded extensions (both from the CWS and from Chrome-Stats), we then collected metadata information from Chrome-Stats.

\subsection{Preprocessing: Extension Unpacking}
\label{subsec:unpacking}

\noindent
Having acquired 165,067 extensions (7,683 from Chrome-Stats and 158,124 from the CWS), we must now unpack them (recall that extensions are \texttt{crx} files; see \Cref{subsec:architecture}) to analyze them. Hence, we rely on a well-known tool, \textsc{DoubleX} {(CCS'21~\cite{DoubleXUnpacker})}, to extract the \texttt{manifest}, content scripts, and service worker of each extension we collected. In doing so, we take extra care to: {\small \textit{(i)}}~remove all extensions that do not have a valid \texttt{manifest}, and {\small \textit{(ii)}}~all extensions that do not have at least one content script or service worker---this is to align our analysis with the scope of our paper (see \Cref{subsec:components}).
After this filtering, we obtain a total of 106,200 extensions: 99,060 \textit{likely benign} (from the CWS) and 7,140 \textit{malicious} (from Chrome-Stats, and flagged as malicious by Google).\footnote{This filtering procedure validates our choice of focusing on \texttt{{\scriptsize manifest}}, \texttt{{\scriptsize service worker}}, and \texttt{{\scriptsize content scripts}}: of the ``known malicious'' extensions collected from Chrome-Stats, only 543 (out of 7,683, i.e., 7\%) have been excluded due to our filtering.}

\subsection{Data Partitioning and Extension Datasets}
\label{subsec:datasets}

\begin{table}[t]
\begin{center}
\resizebox{0.7\columnwidth}{!}{%
\begin{tabular}{l  l  l  l  r}
\toprule
     \textbf{Dataset} & \textbf{Label} & \textbf{Source} & \textbf{Last update} & \textbf{\#Extensions}\\ 
     \midrule
     \seta & benign & CWS & before 2023-01-01 & 63,598\\ 
     \seta & malicious & Chrome-Stats & any & 7,140\\ 
     \midrule
     \setb & unlabeled & CWS & after 2023-01-01 & 35,462\\ 
\bottomrule
\end{tabular}
}
\end{center}
\caption{Summary of \smallseta{} and \smallsetb{}}
\label{tab:datasets}
\end{table}

\noindent
We partition our extensions in two distinct datasets (summarized in \Cref{tab:datasets}), {described below.}

\pseudoparagraph{Rationale}
Among our goals is analysing the effectiveness of detectors of malicious extensions based on supervised ML---i.e, binary classifiers. To develop such classifiers, we need data provided with \textit{ground truth}---i.e., we need a labeled set of benign and malicious extensions.
Unfortunately, there is no existing publicly available dataset of \textit{labeled} browser extensions. As a best-effort strategy, we consider all extensions from the CWS  published or last updated \textit{before Jan. 1, 2023} to be ``benign''.\footnote{We acknowledge that there may be a few malicious extensions that evaded Google's vetting process and that are still in the CWS. However, since the extensions we deem as ``benign'' have been in the CWS for almost a year (we collected our dataset in Nov. 2023, see \Cref{subsec:data-collection}), and malicious extensions take on average one year to be detected~\cite{Hsu2024}, we assume that the fraction of extensions that we (incorrectly) label as benign to be negligible (we further discuss this limitation in \Cref{subsec:limitations}).}
In contrast, we know that the extensions we collected from Chrome-Stats are malicious, since Google engineers analyzed and flagged those extensions as such through their own (proprietary {and closed-source}) checks.
Finally, we consider all extensions published or last updated on the CWS \textit{after Jan. 1, 2023} to be ``unlabeled,'' and we will use these for our open-world assessment. Based on this rationale, we define our two datasets:

\subsubsection{\seta}
\label{subsubsec:datasetA}
We create our ``labeled'' \seta\ by considering all extensions from the CWS whose last update occurred \textit{before January 1, 2023}. Such a cutoff-date leads to \seta\ having 63,598 (out of 99,060) benign extensions from the CWS. 
While there may be some malicious extensions in those, it is impossible to manually vet over 63k extensions; thus, our approach is a \textit{best-effort strategy} to reduce the number of mislabeled extensions while enabling a comprehensive analysis.
Then, we add all 7,140 malicious extensions from Chrome-Stats.  
For our primary investigation, focused on testing our detectors in a ``lab setting'', we further split \seta{} into disjoint training and test sets. We randomly sample 80\% of the benign and 80\% of the malicious extensions to create our training set. The remaining 20\% of \seta{} are used to test our classifiers.

\subsubsection{\setb}
\label{subsubsec:datasetB}
For an open-world evaluation to test our classifiers on ``unknown'' extensions, we create \setb, an \textit{unlabeled} dataset {(with no ground truth)}. \setb{} serves to investigate to what extent there are malicious extensions in the CWS that have not yet been detected. \setb\ contains the remaining 35,462 extensions last updated (or published) \textit{after Jan. 1, 2023}. 
\section{Detectors of Malicious Extensions}
\label{sec:labeled-m-ext-detect}

\noindent
We describe our ML-based approach to detect malicious extensions---our technical contribution.
First, for each extension, we extract a feature vector from its source code and metadata.
Then, we define three classifiers (\code, \meta, and \comb) using our extracted features.
Finally, we evaluate our classifiers on our labeled \seta{} of benign and malicious extensions, and we measure the runtime performance of our approach.

\subsection{Feature Extraction}

\begin{figure}[t]
    \centering
    \includegraphics[width=.8\columnwidth]{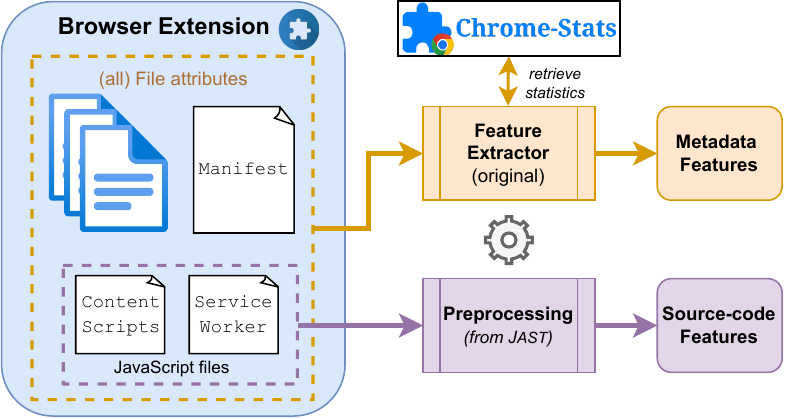}
    \caption{Feature extraction and preprocessing phase of our approach.}
    \label{fig:feature}
\end{figure}

\noindent
Our feature-extraction process is fully automated. A schematic representation of our pipeline is shown in \Cref{fig:feature}.
In what follows, we describe and justify the procedure to extract the feature representation of each browser extension.

\subsubsection{Source Code Features}
\label{subsubsec:source_code_features}

Our ``baseline'' classifier is based entirely on prior work, and analyzes features extracted from extensions' source code. Specifically, we leveraged \jast, a well-known detector of malicious JavaScript code~\cite{JaSt} which has never been applied in the browser-extension context.
\jast{} uses the parser Esprima~\cite{esprima} to build the Abstract Syntax Tree (AST) of an input JavaScript file. Then, it traverses the tree to extract syntactic units and build 4-gram features.
\jast{} leverages the relative frequency of each 4-gram to produce a feature vector of length 2,457. 
In the \jast{} paper~\cite{JaSt}, these features were extracted from JavaScript samples taken from emails or Web pages and used to train a supervised-ML classifier, yielding a detector of malicious JavaScript with 99.5\% accuracy, 0.54\% false negatives, 0.52\% false positives~\cite{JaSt}.
Despite these promising results, however, the JavaScript code of emails or Web pages (i.e., which was the focus of the \jast{} paper~\cite{JaSt}) does not fully align with that of browser extensions (which is the focus of our paper). However, we hypothesize that {\small \textit{(i)}}~the AST of benign and malicious extensions may still differ; and {\small \textit{(ii)}}~extensions may be using code transformation techniques (which leave traces in the AST~\cite{moog2021}), even though this would be a violation of the CWS policies~\cite{be-obf}. Therefore, we use \jast{} to extract AST-based features from the source code of extensions.
Since \jast{} takes a single JavaScript file as input, we concatenated the service worker and content scripts into a single file (see \Cref{subsec:unpacking}).

\subsubsection{Metadata Features}
\label{subsubsec:metadata_features}

To develop our ``original'' classifiers, we extract features purely based on an extension's metadata, i.e., from its \texttt{manifest}, file attributes, and Chrome-Stats' statistics.
In principle, some of these features are inspired by previous work (e.g. permissions~\cite{Hsu2024, Aggarwal2018, Wang2018} or ratings~\cite{Pantelaios2020}). However, the low-level implementation of \textit{76\% of our metadata features is novel}: specifically, we develop a custom feature extractor which we publicly release in our repository~\cite{repository} (we are not aware of any prior work that released a similar tool). Given the novelty of our features, we will validate them while describing them (at a high-level) below.

\begin{table}[t]
    \begin{minipage}{0.45\columnwidth}
        \caption{Top 10 hosts permissions (train set)}
        \label{tab:host}
        \tiny
        \begin{tabular}{ l l l  }
            \toprule
             Rank & Benign extensions & Malicious extensions  \\ 
             \midrule
             1. &  \textless all\_urls\textgreater & *://www.google.com.kh/*  \\ 
             2.  & https://*/* & *://mail.google.com/* \\
             3.  & http://*/* & *://www.google.com.au/* \\
             4.  & *://*/* & *://www.google.us/* \\
             5.  & http://*/ & *://www.google.ca/*\\
             6.  & https://*/ & *://www.google.de/* \\
             7.  & https://ajax.googleapis.com/ & *://www.google.dk/* \\
             8.  & chrome://favicon/ & *://www.google.fr/* \\
             9.  & https://*.1688.com/* & *://www.google.co.jp/* \\
             10.  & https://www.youtube.com/* & *://www.google.nl/* \\
            \bottomrule
        \end{tabular}
    \end{minipage}
    \begin{minipage}{0.45\columnwidth}
        \caption{Top 10 content script matches (train set)}
        \label{tab:content_script}
        \tiny
        \begin{tabular}{ l l l }
            \toprule
            Rank & Benign extensions & Malicious extensions  \\ 
             \midrule
             1. &  \textless all\_urls\textgreater & \textless all\_urls\textgreater  \\ 
             2.  & https://*/* & *://duckduckgo.com/* \\
             3.  & http://*/* & *://gl-search.com/* \\
             4.  & *://*/* & *://redirect.lovelytab.com/* \\
             5.  & https://www.youtube.com/* & *://str-search.com/* \\
             6.  & https://mail.google.com/* & *://search.yahoo.com/search* \\
             7.  & https://twitter.com/* & *://www.google.com/search* \\
             8.  & https://github.com/* & *://www.bing.com/search* \\
             9.  & *://*.youtube.com/* & https://happyhey.com/* \\
             10.  & file://*/* & https://www.happyhey.com/* \\
            \bottomrule
        \end{tabular}
    \end{minipage}
\end{table}

\begin{itemize}[leftmargin=*]
\item \textit{Permissions}: based on Chrome documentation, we collected all 70 permissions an extension can specify in its \texttt{manifest}~\cite{Permissions-list}.
Then, for each extension in our dataset, we analyze its \texttt{manifest} to extract the permissions it specifies. {(We do not consider optional permissions here, because they require user approval during the runtime of an extension.)} This way, each extension has a permission feature vector of length 70 containing booleans: a value of ``true'' indicates that the extension lists the corresponding permission in its \texttt{manifest} and ``false'' otherwise.

\item \textit{Host Permissions}: since host permissions can be arbitrary string-matching patterns, we cannot extract an exhaustive list as before.
Instead, we used our training set (i.e., 80\% of \seta) to extract a subset of the 400 most popular host permissions. Hence, each extension has a boolean feature vector of length 400, where each value indicates whether the extension has the corresponding host permission (``true'') or not (``false'').
We list the top 10 host permissions for benign and malicious extensions in \Cref{tab:host}. Benign extensions tend to use host permissions that can match arbitrary URLs, while malicious extensions list many Google subdomains. These findings align with those in~\cite{Hsu2024}, validating our choice.

\item \textit{Content Script (CS) Matches}: we use the same approach as for the ``Host Permissions'' to extract the top 400 CS matches from the \texttt{manifest} of the extensions in our training set. As before, we get a boolean feature vector of length 400. We list the top 10 CS matches in \Cref{tab:content_script}.
The most popular CS match is ``\textless all\_urls\textgreater'', i.e., CS are injected into any Web pages.
Malicious extensions rather inject their CS in search engine Web pages like ``*://duckduckgo.com/*'' or ``*://gl-search.com/*''. 
This may be due to large clusters of malicious extensions redirecting users' search queries~\cite{Sanchez2022}. Conversely, benign extensions inject their CS into popular pages like YouTube or Twitter. 

\item \textit{Number of CS and Service Worker (SW)}: the last features we obtain from the \texttt{manifest} are the number of CS and SW (or background scripts for Manifest V2) an extension defines, {which are integer values}.

\item \textit{Number of Users}: inspired by~\cite{Jagpal2015}, we use Chrome-Stats to collect the number of users of each extension. 
According to {the} Chrome Web Store Developer Support, the number of active users of an extension is ``the number of Chromes with the extension installed that are active and checking in to [their] update servers over the previous seven days only, not for all time. It is not equal to the sum of historic installs minus the sum of historic uninstalls''~\cite{Hsu2024}.

\item \textit{Number of Ratings and Average Rating Score}: users can rate extensions with {1--5} stars on the CWS. We use the total number of ratings received by an extension, as well as an extension's average rating, as features (for the latter, the value is 0 if an extension has no rating). 

\begin{table}[!t]
    \footnotesize
    \caption{Most common keywords among the extensions in the \textit{training set}}
    \label{tab:keywords}
    \begin{subtable}[t]{0.3\textwidth}
        \resizebox{1\columnwidth}{!}{%
            \begin{tabular}{ l l l  }
                \toprule
                Rank & Benign Ext. & Malicious Ext.  \\ 
                \midrule
                1. &  EXTENSION & GOOD  \\ 
                2.  & WORK & LOVE \\
                3.  & GREAT & EXTENSION \\
                4.  & GOOD & LIKE \\
                5.  & THANK & GREAT \\
                6.  & NOT WORK & WORK \\
                7.  & USE & THANK \\
                8.  & LIKE & USE \\
                9.  & LOVE & NICE \\
                10.  & TIME & TIME \\
                \bottomrule
            \end{tabular}
        }
        \caption{Top 10 review keywords}
        \label{tab:review}
    \end{subtable}
    \begin{subtable}[t]{0.3\textwidth}
        \resizebox{1\columnwidth}{!}{%
            \begin{tabular}{ l l l  }
                \toprule
                Rank & Benign Ext. & Malicious Ext.  \\ 
                \midrule
                1. &  EXTENSION & EXTENSION  \\ 
                2.  & CLICK & NEW \\
                3.  & PAGE & TAB \\
                4.  & CHROME & TIME \\
                5.  & USE & WALLPAPER \\
                6.  & ADD & FAVORITE \\
                7.  & NEW & THEME \\
                8.  & TIME & HIGH \\
                9.  & WEBSITE & BACKGROUND \\
                10.  & BROWSER & FEATURE \\
                \bottomrule
            \end{tabular}
        }
        \caption{Top 10 summary keywords}
        \label{tab:summary}
    \end{subtable}
    \begin{subtable}[t]{0.3\textwidth}
        \resizebox{1\columnwidth}{!}{%
            \begin{tabular}{ l l l  }
                \toprule
                 Rank & Benign Ext. & Malicious Ext.  \\ 
                 \midrule
                 1. &  EXTENSION & NEW  \\ 
                 2.  & CHROME & TAB \\
                 3.  & PAGE & WALLPAPER \\
                 4.  & TAB & HD \\
                 5.  & ADD & CHROME \\
                 6.  & WEB & THEME \\
                 7.  & BROWSER & WALLPAPERS \\
                 8.  & WEBSITE & USEFUL \\
                 9.  & ALLOW & UTILITY \\
                 10.  & CLICK & LOT \\
                \bottomrule
            \end{tabular}
        }
        \caption{Top 10 description keywords}
        \label{tab:description}
    \end{subtable}
\end{table}

\item \textit{Description, Summary, and Review Keywords}: we extract features from the description, summary, and reviews of an extension.
The description is a short text describing the functionality of an extension.
The summary is typically a longer article that contains more information about the behavior of an extension.
The reviews are written by users on the CWS to evaluate an extension.
In all three cases, we need to process these free texts before being able to extract features.
To this end, we use spaCy \cite{spacy}, a natural language processing tool to {\small \textit{(i)}}~remove stop words and punctuation; and {\small \textit{(ii)}}~obtain the basic form of words, e.g., this turns an ``is'' into a ``be''. Finally, we split each text into a set of single words and remove duplicates per text.
This way, we get three sets of words for each extension (for description, summary, and reviews).
As before, we compute the top 400 words for description, summary, and reviews; and we check for their presence for each extension we analyze.
Tables \ref{tab:keywords} show the top 10 description, summary, and review keywords.
Interestingly, the keywords ``NEW'', ``TAB'', and ``WALLPAPER'' are frequently used in the description and summary of malicious extensions. This is probably due to the cluster of ``new tab wallpaper'' extensions, which change the appearance and functionality of a newly opened tab~\cite{Pantelaios2020} and sometimes also show advertisements or contain malware~\cite{Hsu2024}.

\item \textit{Same Developer Count}: we consider the number of extensions by the same developer (i.e., same username and email address). We obtain this number by using Chrome-Stats' advanced search functionality for each extension, as also done in~\cite{Hsu2024}.

\item \textit{File Count, \texttt{CRX} Size, JavaScript File Count, JavaScript Size}: additionally, we use file attributes as features. We consider the size of a \texttt{CRX} file in bytes, the number of files and JavaScript files inside of an extension, and the size of those JavaScript files in bytes.

\item \textit{Related Permissions}: for each extension, we extract the permissions of the first four similar extensions recommended by Google. (We manually verified that the recommendation list is stable over time---as also done by~\cite{Kim2023}.)
We then compare the permissions specified in an extension's \texttt{manifest} with those of the four recommended extensions.
For each extension, we have a feature vector of length 70 (the maximum {number of} declarable permissions). Initially, all the values are set to 0. If the analyzed extension declares a permission that is not listed by $n$ ($n \in \llbracket 1, 4 \rrbracket$) recommended extension(s), then we decrease the value by $n$.
\end{itemize}

\summary{Remark.}{In devising our metadata features, we use the extensions in the \textit{training set}, i.e., we follow the recommendations by Arp et al.~\cite{arp2022and} and do not commit the ``data snooping'' crime---which would unfairly increase the performance of our classifiers on the test set.}

\subsubsection{Features Summary.}
\label{subsubsec:summary}
\Cref{tab:features} summarizes all the extracted features (metadata and source code), including the length of each feature vector and how we extracted each feature. The low-level technical details that enable complete reproduction of our feature-extraction procedure can be found in our repository~\cite{repository}.

\begin{table}[ht]
    \vspace{-2mm}
    \footnotesize
    \begin{center}
        \resizebox{0.65\columnwidth}{!}{%
            \begin{tabular}{ l r l }
                \toprule
                Feature name & Feature length & Feature origin \\ 
                \midrule
                Permissions & 70 & \texttt{{\footnotesize manifest}} \\ 
                Host Permissions & 400 & \texttt{{\footnotesize manifest}}  \\ 
                Content Script Matches & 400 & \texttt{{\footnotesize manifest}} \\
                Number of Content Scripts & 1 & \texttt{{\footnotesize manifest}} \\
                Number of Service Workers & 1 & \texttt{{\footnotesize manifest}} \\
                Number of Users & 1 & Chrome-Stats \\
                Average Rating Score & 1 & Chrome-Stats \\
                Number of Ratings & 1 & Chrome-Stats \\
                Description Keywords & 400 & Chrome-Stats \\
                Summary Keywords & 400 & Chrome-Stats  \\
                Review Keywords & 400 & Chrome-Stats  \\
                Same Developer Count & 1 & Chrome-Stats \\
                \texttt{CRX} Size & 1 & File attributes  \\
                File Count & 1 & File attributes  \\
                JavaScript File Count & 1 & File attributes \\
                JavaScript Size & 1 & File attribute  \\
                Related Permissions & 70 & \texttt{{\footnotesize manifest}} + Chrome-Stats \\
                \midrule
                Source Code (AST) & 2,457 & \jast~\cite{JaSt} \\
                \bottomrule
            \end{tabular}
        }
    \end{center}
    
    \caption{{Summary of the extracted features for each extension}}
    \label{tab:features}
    \vspace{-3mm}
\end{table}

We mention that, during the extraction process of the source-code features, the generation of the AST by \jast{} led to syntax errors (raised by Esprima) for 7,448 out of 106,200 extensions. In contrast, we encountered no errors in the generation of the metadata-related features. Despite this small discrepancy (explicitly quantified, for both \seta{} and \setb{}, in Table~\ref{tab:parsing-issues}), we will not exclude extensions for which we could obtain only their metadata-related features. We will discuss the consequences of such a design choice in~\Cref{subsec:limitations}.

\begin{table}[ht]
    \footnotesize
    \begin{center}
        \begin{tabular}{l  r  r }
            \toprule
            & Metadata features & Source code features \\ 
            \midrule
            \#Extensions & 106,200 & 98,752 \\ 
            - \seta & 70,738 & 68,748 \\ 
            - \setb & 35,462 & 30,004 \\ 
            \bottomrule
        \end{tabular}
    \end{center}
    \caption{Extensions with metadata and source code features extracted}
    \label{tab:parsing-issues}
\end{table}

\begin{figure}[t]
    \centering
    \includegraphics[width=.95\columnwidth]{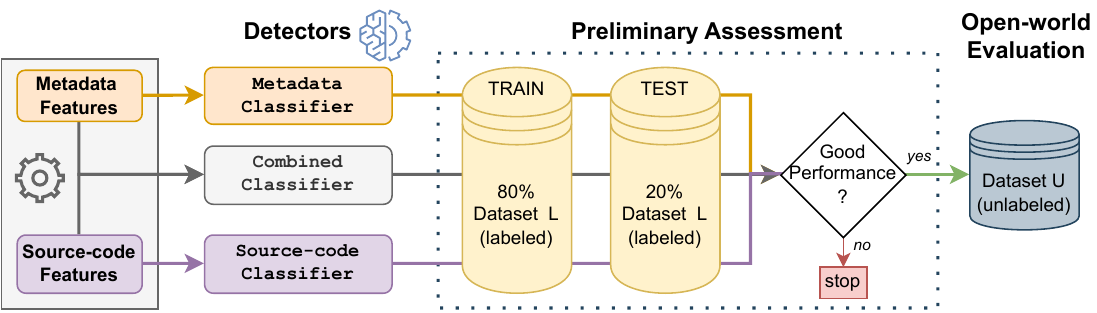}
    \caption{{Workflow: iif our detectors exhibit good performance on \smallseta{}, we use them on \smallsetb{}}}
    \label{fig:workflow}
\end{figure}

\subsection{Classifier Development}
\label{subsec:detector_def}

\noindent
For our large-scale assessments of supervised-ML methods, we develop three (binary) classifiers:
\begin{itemize}[leftmargin=*]
    \item \textbf{\code}: this ``baseline'' classifier (inspired by \jast{}~\cite{JaSt}) leverages only features from the extensions' source code (see \Cref{subsubsec:source_code_features});
    \item \textbf{\meta}: this ``original'' classifier uses only the extensions' metadata features (see \Cref{subsubsec:metadata_features});
    \item \textbf{\comb}: this classifier uses both metadata and source code features---thereby enhancing \jast{}~\cite{JaSt} with our custom-defined metadata features.
\end{itemize}

\noindent
As shown in \Cref{fig:workflow}, we use \seta{} to train each classifier (i.e., a detector) and preliminary assess its performance in a ``lab setting''. Then, if the performance is good, we will use \setb{} to evaluate the classifier in an open-world setting (covered in \Cref{sec:open}). In the remainder, we use the terms ``classifier'' and ``detector'' interchangeably; a ``positive'' denotes a ``malicious'' sample.

To develop our classifiers, we empirically evaluated several classification algorithms available in scikit-learn {(e.g., gradient boosting, naive bayes, and even deep neural networks)} and eventually chose random forest~\cite{RandomForest}, which provides the best detection performance (after cross-validation on the training set). This result is in line with prior work~\cite{JaSt, xxjstap}.\footnote{{Indeed, tree-based algorithms are known~\cite{grinsztajn2022tree} to be superior to deep learning ones for analyzing (security-related) tabular data---both in terms of detection performance and operational runtime~\cite{apruzzese2022sok}. {In our case, a deep neural network (using a multi-layer perceptron)} took 2.5x the training time and classified everything as benign.}} 

Given that the benign extensions in the training set vastly outnumber malicious extensions (i.e., the benign:malicious ratio is almost 9:1) we set the class weight to ``balanced'', which adjusts the weights inversely proportional to the number of samples in each class in the training set. 
To select the optimal set of hyperparameters, we performed 5-fold cross validation on the training set. We empirically inferred that 300 decision trees provide the best trade-off between detection accuracy and runtime performance for all three classifiers.
As an original design choice, our classifiers seek to minimize both the false-positive and false-negative rates (instead of, e.g., focusing on minimizing just one of these two metrics\footnote{We will, however, also consider detectors that minimize the false-positive rate (see Section~\ref{subsec:falsepositive}).}). We do this by leveraging the Youden's $J$ statistic~\cite{YoudenJ1950, PowersJ2011Youden}:
	$$ J = \text{sensitivity} + \text{specificity} - 1 = TPR - FPR $$
    
\begin{figure*}[!htbp]
    \centering
    \begin{subfigure}{0.28\textwidth}
        \centering
        \includegraphics[width=\columnwidth]{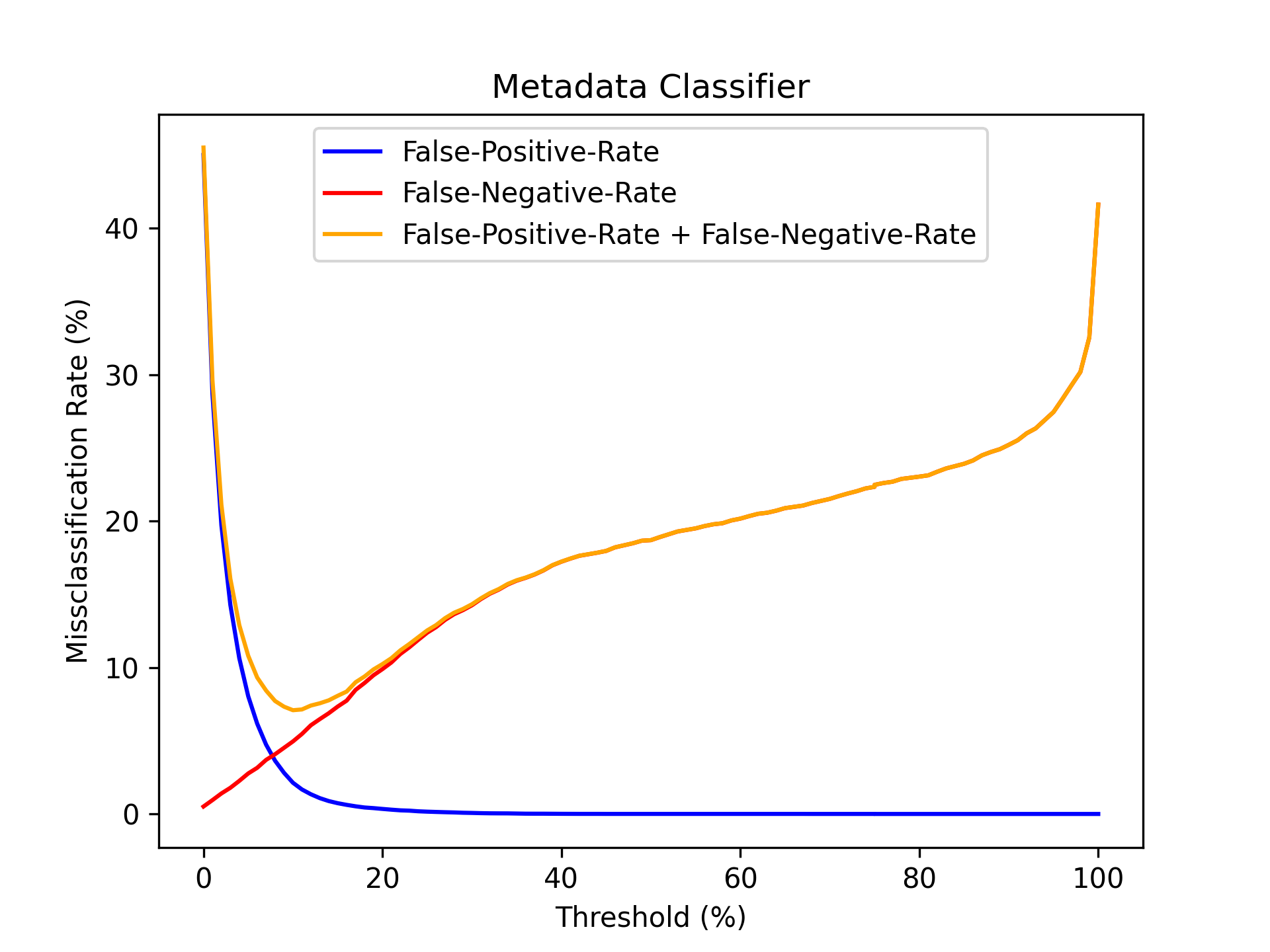}
        
        \caption{\smallmeta}
        \label{fig:threshold_meta}
    \end{subfigure}
    \begin{subfigure}{0.31\textwidth}
        \centering
        \includegraphics[width=0.9\columnwidth]{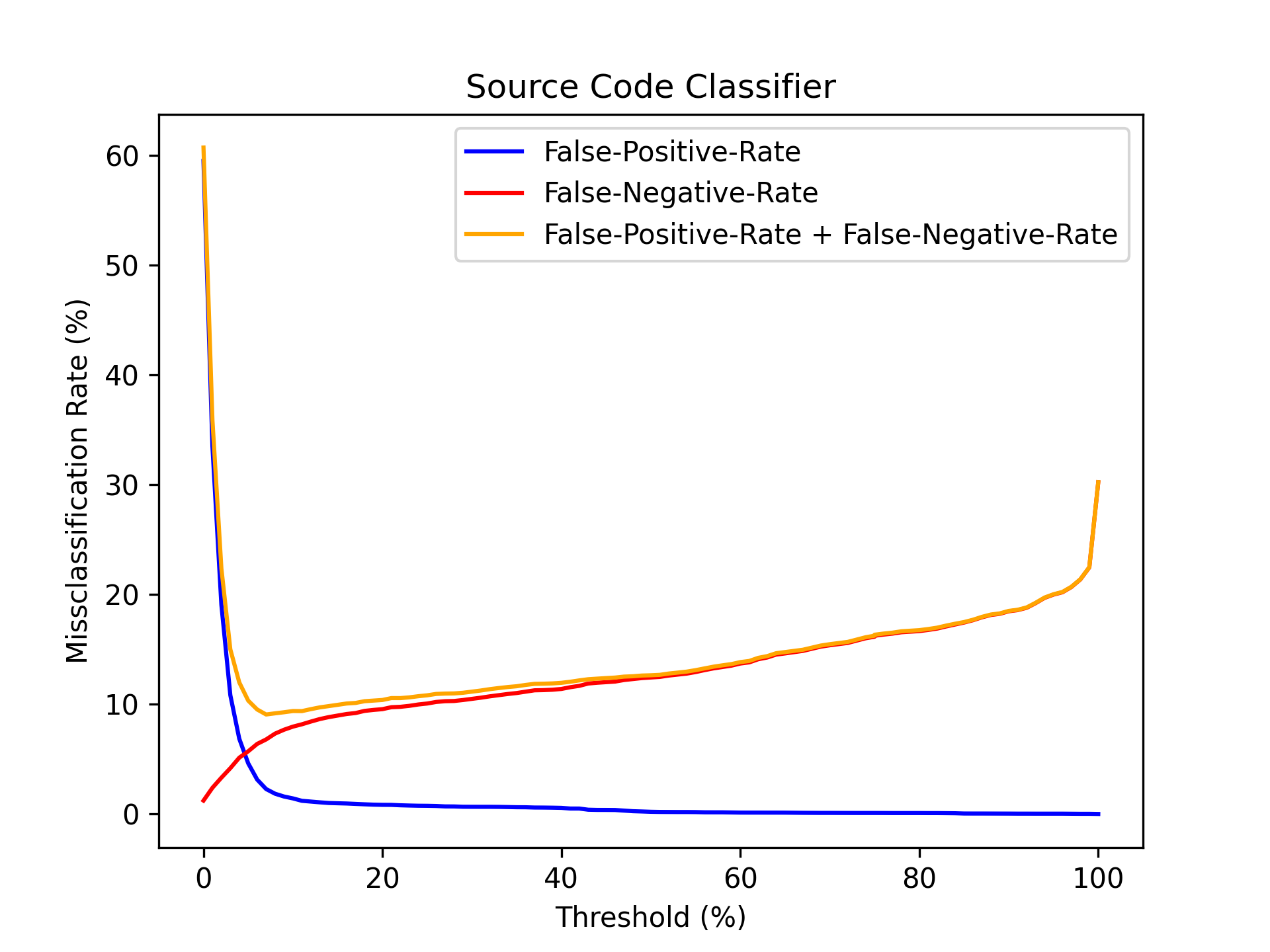}
        
        \caption{\smallcode}
        \label{fig:threshold_jast}
    \end{subfigure}
    \begin{subfigure}{0.31\textwidth}
        \centering
        \includegraphics[width=0.9\columnwidth]{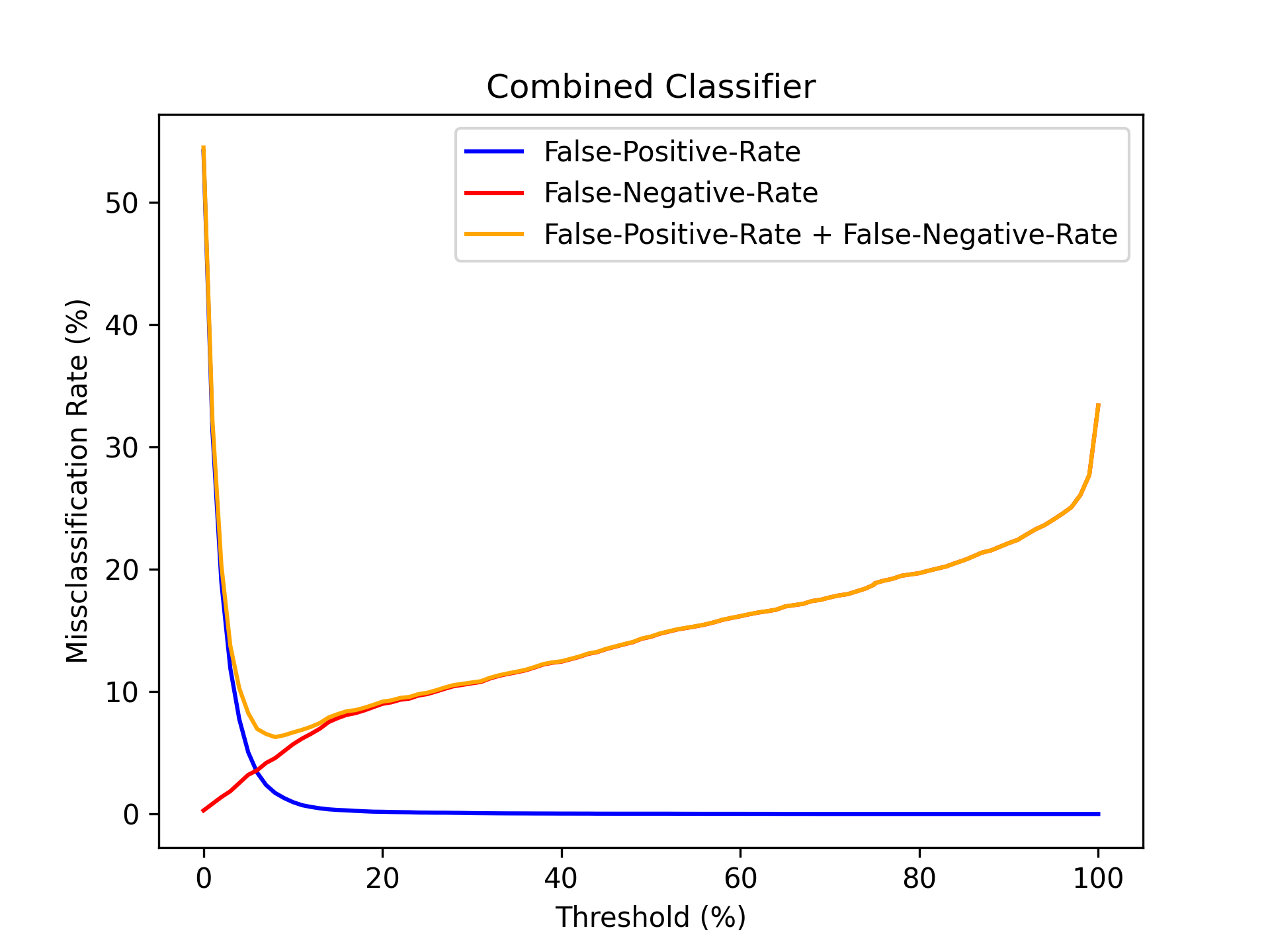}
        
        \caption{\smallcomb}
        \label{fig:theshold_combined}
    \end{subfigure}
    \caption{Evolution of the false-positive and false-negative rates depending on the threshold value}
    \label{fig:threshold}
\end{figure*}

We empirically assessed which threshold value achieves the lowest cumulative sum of the false-negative and false-positive rates; and we averaged the value across the 5 cross-validation runs (always performed on the training set) for each of the three classifiers.
We display the results in \Cref{fig:threshold}. The optimal threshold value for the \meta{} is 11.6\%; i.e., if 11.6\% of the decision trees predict that an extension is malicious, it will be flagged as malicious. For the \code, the threshold is 8.8\% and 9.2\% for the \comb.

\subsection{Evaluation (before {real-world} deployment)}
\label{subsec:classification_results}

\begin{figure}[t]
    \centering
    \includegraphics[width=\columnwidth]{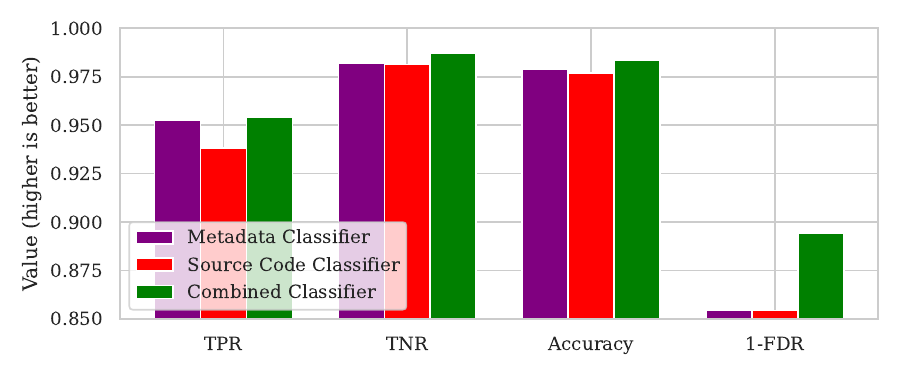}
    \vspace{-7mm}
    \caption{Accuracy, TPR, TNR, and precision (1-FDR) of our classifiers}
    \label{fig:perf-clf}
\end{figure}

\noindent
We present in~\Cref{fig:perf-clf} the results of our three classifiers (\texttt{Metadata}, \texttt{Source code}, and \comb) on the test set of our labeled \seta{}. Let us discuss these---very encouraging---results.

\subsubsection{\code}
\label{subsection:sourcecode_results}

First, we consider our ``baseline'' classifier solely based on source-code features from {\jast}~\cite{JaSt}.
This detector achieves a high accuracy, with 97.69\% of the extensions being correctly classified (\Cref{fig:perf-clf}). 
In addition, we have a high-true negative rate of 98.14\% (12,107\,/\,12,337) and true-positive rate of 93.83\% (1,324\,/\,1,411). These results reveal that our \code{} (which analyzes the JavaScript code of browser extensions) has a slightly inferior performance to \jast{} (which was designed for detecting malicious JavaScript in Web pages or emails). Indeed, the results in~\cite{JaSt} show an accuracy of 99.5\%, true-positive rate of 99.46\%, and true-negative rate of 99.48\%. This means that, while the AST of benign and malicious extensions differ, the differences are less pronounced than on benign vs.\ malicious Web pages {or emails}. Nevertheless, our results suggest that using AST-related features can be an effective way to detect malicious browser extensions---representing a solid baseline, and validating our hypotheses.

\subsubsection{\meta}
\label{subsection:metadata_results}

Next, we focus on our ``original'' classifier.
We find that malicious extensions can be detected with 97.88\% accuracy by using only metadata information (\Cref{fig:perf-clf}).
In particular, 12,488 (out of 12,720) benign extensions are correctly classified as benign; this represents a high true-negative rate (TNR) of 98.18\%. We also have a high true-positive rate (TPR) of 95.24\%, with 1,360 (out of 1,428) malicious extensions correctly identified. However, due to the class imbalance, we have a false-discovery rate (FDR) of 14.57\%, i.e., of the 1,592 extensions flagged as malicious, 232 are actually benign---suggesting that there is room for improvement.
Nonetheless, altogether, these results suggest that the \meta{} has similar performance to the \code. This is interesting given that the malicious functionality of extensions is reflected in their \textit{source code}, and yet it is still possible to detect such malicious extensions by only using metadata-related information (a finding that contrasts what was suggested in some prior work~\cite{ban2019}).

\begin{figure}[t]
    \centering
    \begin{subfigure}{0.45\textwidth}
        \centering
        \includegraphics[width=\columnwidth]{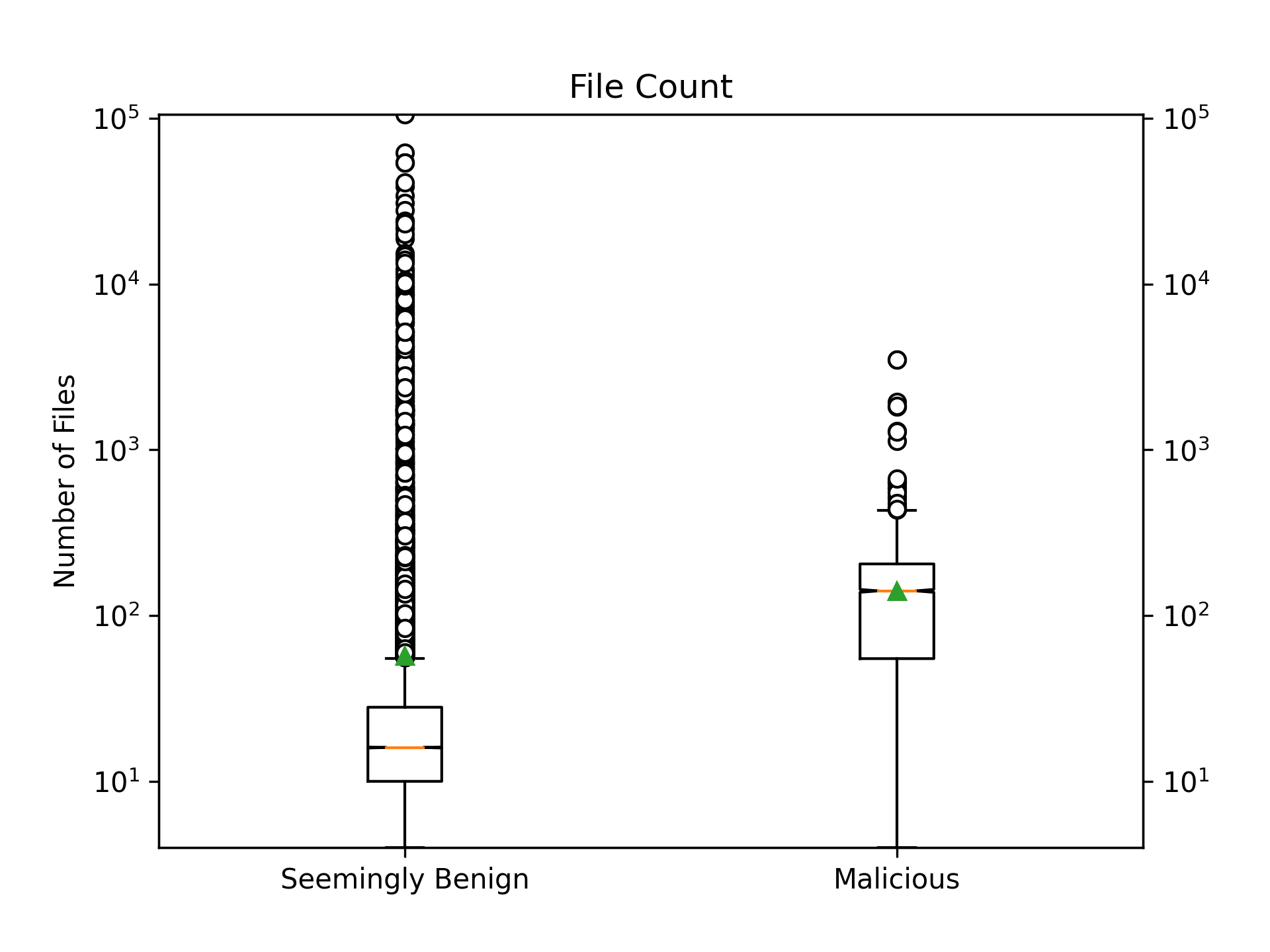}
        \caption{Number of files}
        \label{fig:file_count}
    \end{subfigure}
    \begin{subfigure}{0.45\textwidth}
        \centering
        \includegraphics[width=\columnwidth]{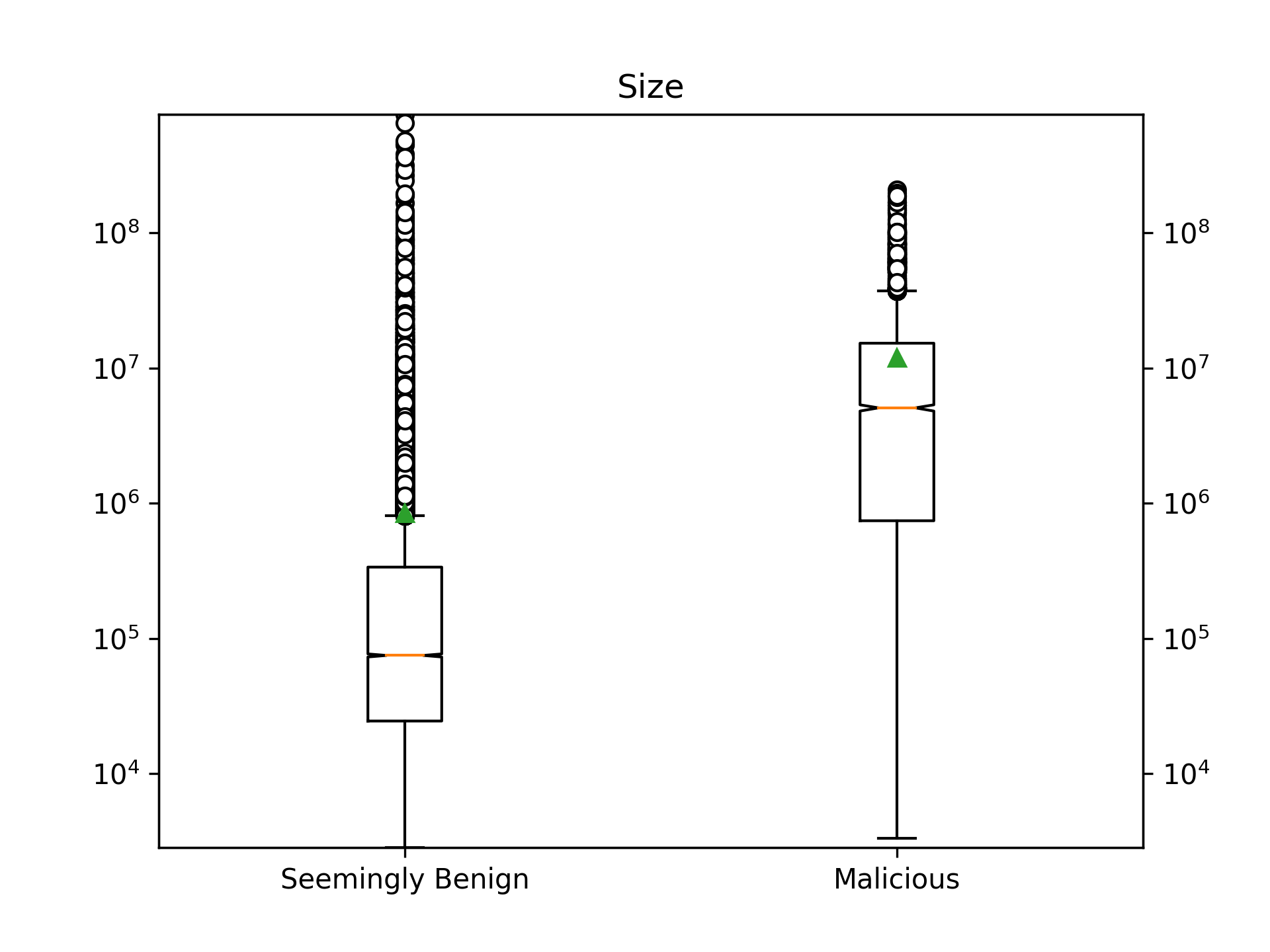}
        \caption{Size of an extension in bytes}
        \label{fig:size}
    \end{subfigure}
    \caption{{Distribution of two features for benign \& malicious extensions}}
\end{figure}

To better understand the logic behind the predictions of our classifier, we {\textit{studied the importance of each feature}}, by leveraging the mean decrease in impurity~\cite{MeanDecrease}.
The most important feature is the number of files included in an extension (``File Count'').
As shown in \Cref{fig:file_count}, benign extensions include a median (orange line) of 16 files vs.\ 141 for malicious extensions. Similarly, benign extensions contain 57.23 files on average (green triangle) and malicious extensions 140.18.
We assume that malicious extensions may split their source code into multiple files to make their analysis more difficult, which we empirically confirmed in~\Cref{subsection:manual-analysis}, during our manual analyses.
Besides summary and description keywords related to New Tab Extensions (e.g., ``TAB'', ``WALLPAPER'', or ``NEW''), the fourth most important feature is related to the size of an extension.
\Cref{fig:size} shows the distribution of the ``Size'' among benign and malicious extensions.
Malicious extensions are larger (mean of 11.94 MB) than benign extensions (0.85 MB). In particular, we observed that malicious extensions contain obfuscated code and dead code to make their analysis more difficult.
The fifth most important feature is the ``Same Developer Count'', which indicates how many extensions were published on the CWS by the same developer.
While a developer with only benign extensions publishes on average 2.7 (median of 1) extensions, a malicious developer publishes 337.58 extensions on the CWS on average (median of 18).
This finding shows that developers who upload malicious extensions tend to be ``repeated offenders'', and hence \textit{should be closely monitored by Google} for future uploads.\footnote{Moreover, ``repeated abuse'' should result in the closure of the account~\cite{repeatabuse}, but such accounts were still able to publish hundreds of extensions. Nonetheless, once Google detects (and removes) malicious browser extensions by a given developer, other (malicious) extensions from the same developer can stay in the CWS for months before being removed by Google~\cite{Hsu2024}.}

\subsubsection{\comb}
\label{subsection:combined_results}
Finally, we investigate to what extent the combination of metadata and source-code features improves the detection of malicious browser extensions.
As shown in \Cref{fig:perf-clf}, this detector outperforms the other two: it accurately classifies 98.37\% of the extensions (13,524 out of 13,748). 
It also has a higher true-negative rate (98.71\%) and higher true-positive rate (95.39\%) than the other two classifiers.
In addition, we have a false-discovery rate of 10.56\% (159 false positives out of 1,505 extensions classified as malicious), which is lower than before.

\subsubsection{Runtime Performance}
\label{subsubsec:runtime}
We evaluated the operational runtime of our approach end-to-end, i.e., feature extraction and classification, plus training. We carry out our experiments on a server mounting four AMD EPYC 7H12 CPUs (each having 64 cores and 128 threads) and 2 TB RAM; we do not use GPUs. We parallelize our experiments to use all cores available. Overall, given an extension, 0.112s are required to extract the metadata features, and 0.673s for the source code features. To train our classifiers (on 80\% of \seta), we need: 25s for \meta{}, 27s for \code{}, and 44s for \comb{}. The classification time (per extension) is: 1.88ms for \meta{}, 2.81ms for \code{}, 4.69ms for \comb{}.
So, on our testbed, an extension can be analyzed end-to-end in less than 1~second with our classifiers.\footnote{We also repeated the ML-runtime experiments (training and testing) on a different setup: an AMD Ryzen 5800X3D @ 4.5GHz (with 8 cores and 16 threads) having 32GB of RAM. To simulate a worst-case scenario, we forced the usage of only one thread (which should significantly impact the training time). Over five independent runs, it takes an average of 140s to train the \texttt{\scriptsize Combined classifier}, and the test phase requires 2.2s, i.e., around 0.16ms per sample.}

\summary{Takeaway}{Our three classifiers can detect malicious extensions with high accuracy ($>$97.69\%) as well as low false-positive ($<$1.86\%) and false-negative rates ($<$6.17\%).
The \comb{} achieves the best results given the combination of both metadata and source code information.
Moreover, our detectors can analyze an extension end-to-end in less than 1~second. 
Altogether, these results show that our detectors can \textit{in theory} be used to automatically flag malicious extensions whenever a new extension is uploaded to the CWS.}

\section{Analyzing Extensions in the Open World}
\label{sec:open}

\noindent
Our experiments in \Cref{sec:labeled-m-ext-detect} showed that our classifiers perform well to detect malicious extensions {in a ``lab setting}''.
Hence, we now assess our detectors' performance in an open-world setting---i.e., to scrutinize the presence of malicious extensions in our unlabeled \setb.

\subsection{Classification of Unlabeled Extensions}
\label{subsec:classification_unlabeled}

\pseudoparagraph{Classifier Predictions}
As described in \Cref{subsubsec:datasetB}, \setb{} has extensions that were published or received an update on the CWS \textit{after Jan. 1, 2023}. The findings of prior work suggest~\cite{Hsu2024} 
that this dataset may contain some malicious extensions that have not yet been detected by Google nor security researchers.
Therefore, we analyze the 35,462 extensions from \setb{} with our three detectors.
\Cref{tab:classifier_b} summarizes the predictions of our classifiers.
Surprisingly, between 2,638 (\code) and 3,793 (\meta) extensions are flagged as malicious. \textit{This represents almost 10\% of our dataset}.
Even if we take the false-discovery rate of our classifiers into account (between 10.56--14.80\%, see \Cref{subsec:classification_results}), we would still flag between 2,248 and 3,240 extensions as malicious with each classifier.

\begin{table}[t]
    \footnotesize
    \begin{center}
    \resizebox{0.6\columnwidth}{!}{%
        \begin{tabular}{ l r r|r }
            \toprule
             & Benign & Malicious & Total \\
             \midrule
             \meta & 31,669 & 3,793 & 35,462\\
             \code & 27,366 & 2,638 & 30,004 \\
             \comb & 27,039 & 2,965 & 30,004 \\
            \bottomrule
        \end{tabular}
    }
    \end{center}
    \caption{Open-world classification results on \smallsetb, which require further analyses. \textmd{N.b.: we cannot extract source-code features for some extensions due to parsing errors in their source files (\Cref{subsubsec:source_code_features})}}
    \label{tab:classifier_b}
\end{table}

\pseudoparagraph{Intersection of Flagged Extensions}
We have reason to believe that our results overestimate the number of malicious extensions---but confirming this hypothesis requires manual review of each extension flagged as malicious. However, it is not feasible to manually check \textit{thousands} of extensions. Hence, as a best-effort strategy, and  to increase the likelihood that a reported extension is truly malicious, we focus on extensions that are \textit{classified as malicious by all three detectors}.\footnote{To justify this choice, we carry out this ``intersectional'' experiment on our labeled \smallseta. Overall, 1,321 extensions were classified as malicious by all three classifiers, and only 25 were false positives. This corresponds to a false-discovery rate of 1.89\%, which is significantly lower than the 10.56--14.80\% we found in \Cref{subsec:classification_results}. Moreover, the false-negative rate is only slightly impacted (8.15\% vs 4.61--6.17\%).} 
In this way, we are using our classifiers as an ensemble architecture.
On our unlabeled \setb, there are 1,131 extensions that are classified as malicious by the ensemble. We will further scrutinize these extensions to identify which ones are truly malicious. To this end, we first consider using operational products, and then resort to manual analyses.

\subsection{Effectiveness of Existing Security Tools} 
\label{subsec:existing_tools}
\noindent
We first investigate if well-known security solutions can be used to provide meaningful analyses for the sake of ascertaining the ``maliciousness'' of browser extensions.

\pseudoparagraph{VirusTotal}
As a starting point, we consider VirusTotal, which is well known in security research~\cite{zhu2020measuring}. As of {Oct.} 2024, VirusTotal uses over 70 different security vendors (including Kaspersky, McAfee, and Avira) to analyze files, domains, IPs, and URLs to detect malware. However, the results of VirusTotal may be inconsistent across engines~\cite{zhu2020measuring}.
Moreover, a study carried out in 2017~\cite{DeKoven2017} found that VirusTotal was able to detect only 5 out of 9k malicious extensions. Therefore, to estimate the reliability of VirusTotal to detect malicious browser extensions ``today,'' we verified \textit{if VirusTotal was able to detect known malicious extensions}.
{[\underline{Results}]}~We sent {our set} of 7,140 \textit{known} malicious extensions ({recall that those extensions were flagged as malicious by Google itself}) from \seta{} to the VirusTotal API.
{Of those, only 293} were flagged as malicious by at least one engine. More specifically: {120 were flagged by one vendor, 104 by two, 26 by three, 8 by four, 3 by five, 8 by 6--10 engines, 15 by 11--20, and 9 by 21--30.}
According to prior work~\cite{zhu2020measuring}, it is recommended to consider at least more than one engine to derive sound conclusions: this would lead to VirusTotal being able to detect only {173} malicious extensions out of the {7,140} we submitted---i.e., a \textit{false-negative rate of {97.6}\%}.
It is unclear why VirusTotal is not able to detect the huge majority of known malicious extensions identified by Google.
So, this ``negative result'' (on recent data) led us to look for alternative solutions to analyze the 1,131 extensions from \setb{} flagged as malicious by our classifiers. Nevertheless, the fact that commercial security engines cannot detect \textit{known} malicious extensions is a sign that more work is needed in this area.

\pseudoparagraph{CRXcavator}
This tool, developed by Duo's Corporate Security Engineering team~\cite{CRXcavator}, provides a \textit{Risk Score} that can be used to estimate the ``maliciousness'' of a given extension. 
It leverages factors, such as an extension permissions, inclusion of vulnerable third-party JavaScript libraries, 
or missing details from the associated description on the CWS to assign ``points of risk'' to each extension. The resulting Risk Score is the sum of all the points of risk: lower (higher) values denote smaller (greater) chances that an extension presents security risks. 
We hence retrieve the corresponding scores for the 1,131 extensions flagged as malicious by our ensemble on \setb{}. Moreover, to provide a more comprehensive overview, we also retrieve the risk scores for the remaining 34,331 ``unlabeled'' extensions of \setb{}, i.e., those that have not been flagged as malicious by the ensemble (and which we consider as less likely to be malicious). [\underline{Results}]~\Cref{tab:riskscore}~shows the percentage of extensions with a given Risk Score.
We see that 73.42\% of flagged extensions have a ``moderate'' risk score, ranging from 400--599; in contrast, only 51.02\% of the remaining extensions fall within this range. However, by focusing on extensions with the highest risk score (above 600), we see that the remainder of \setb{} (which is less likely to be malicious) has more candidates than the extensions in the flagged set (which are more likely to be malicious). This is true both in absolute (1,431 vs 34) and relative (4.17\% vs 3\%) terms. Hence, the Risk Score is not a metric we can rely on to assess the maliciousness of the extensions flagged as malicious by our ensemble.

\begin{table}[!t]
    \vspace{0.8mm}
    \begin{center}
        \resizebox{0.9\columnwidth}{!}{%
            \begin{tabular}{ l | r r r r r r | r}
                \hline
                & \cellcolor{green} \textless 300 & \cellcolor{green} 300-399 & \cellcolor{orange} 400-499 & \cellcolor{orange} 500-599 & \cellcolor{red} 600-699 & \cellcolor{red} \textgreater 699 & \#Ext \\
                \hline
                Flagged by all 3 classifiers & 0.46 & 23.12 & 41.71 & 31.71 & 1.90 & 1.10 & 1,131 \\
                \setb{} minus flagged & 0.34 & 44.47 & 44.52 & 6.50 & 2.17 & 2.00 & 34,331\\
                \hline
            \end{tabular}
        }
    \end{center}
    \caption{Risk Score (CRXcavator), in percent of extensions flagged as malicious by the 3 classifiers and remaining \smallsetb{}}
    \label{tab:riskscore}
\end{table}

\begin{table}[!t]
    \begin{center}
        \resizebox{0.8\columnwidth}{!}{%
            \begin{tabular}{ l | r r r r r | r }
                \hline
                & \cellcolor{green} 0 & \cellcolor{green} 1 & \cellcolor{orange} 2 & \cellcolor{red} 3 & \cellcolor{red} 4 & \#Ext\\
                \hline
                Flagged by all 3 classifiers & 11.09 & 34.61 & 34.25 & 11.98 & 8.07 & 1,131\\
                \setb{} minus flagged & 12.32 & 6.18 & 53.96 & 26.13 & 1.41 & 34,331\\
                \hline
            \end{tabular}
        }
    \end{center}
    \caption{Risk Likelihood (Chrome-Stats), in percent of extensions flagged as malicious by the 3 classifiers and remaining \smallsetb{}}
    \label{tab:risklikelihood}
\end{table}

\pseudoparagraph{Chrome-Stats}
As a final attempt, we consider another risk-based metric. Indeed, Chrome-Stats provides the \textit{Risk Likelihood}, which measures the risk that an extension is malicious based on, e.g., the reputation of its developer and of the extension on the CWS, or how long it has been available. The risk is an integer between 0 (low) and 4 (high). To align this assessment with the one for CRXcavator, we retrieve the Risk Likelihood for all extensions in \setb{}. [\underline{Results}]~\Cref{tab:risklikelihood} shows the percentage of extensions with a given Risk Likelihood score.
By focusing on the lower scores, we observe that almost half (45.7\%) of the 1,131 extensions flagged as malicious by our ensemble have a low risk score of 1 or 0; and this percentage raises to 80\% if we also include a ``medium'' risk score of 2. This may suggest that over 80\% of our 1,131 extensions are not truly malicious. However, by focusing on the higher scores (denoting a higher risk of maliciousness), we see an intriguing result: while 20\% (231) of the extensions flagged as malicious by our ensemble have a score of 3 or 4, this holds true for 27.54\% (9,454) of the remaining extensions in \setb{}, which are less likely to be malicious. Moreover, by focusing on the \textit{absolute number} of extensions with a score of 4, a total of 575 in \setb{} achieve the highest risk likelihood: if we deem these as malicious, it would mean that the CWS would have hosted 575 malicious extensions (on Nov. 16th, 2023) that were reported as malicious by Chrome-Stats---which we {consider as an unrealistically high number}.
Hence, we believe that the Risk Likelihood metric cannot be used {either to verify if our flagged extensions are truly malicious or not.}

\summary{Takeaway}{Commercial products like VirusTotal perform poorly and cannot detect most extensions \textit{known} to be malicious. Risk scores provided by Chrome-Stats and CRXcavator also provide contrasting results.
Hence, these tools do not enable us to {assess if the extensions flagged by our three detectors are indeed malicious.}}

\subsection{Manual Analysis}
\label{subsection:manual-analysis}

\noindent
Since we cannot rely on existing tools to identify malicious extensions, we manually reviewed a subset (200, i.e., 18\%) of the 1,131 extensions flagged as malicious by our ensemble of three~classifiers.

\pseudoparagraph{Methodology} To carry out a feasible manual analysis, we {\small \textit{(i)}}~group our extensions into clusters, {similarly to~\cite{Pantelaios2020}}, and then {\small \textit{(ii)}}~analyze a few extensions for each cluster. To this end, and inspired by prior work~\cite{Hsu2024}, we
used ssdeep~\cite{fuzzy_hash} to compute the fuzzy hash of the concatenated (content scripts, service worker) files, and we grouped extensions with a similarity score $>$90 in the same cluster.
We could group 732 (out of 1,131) extensions into 93 clusters, having a minimum cluster size of 2 extensions, maximum of 157, and average of 8.
We ignored the remaining extensions for our manual analysis, since they could not be grouped with others.
Then, for each cluster, we inspected a subset of the extensions by reviewing their code. {When necessary}, we used js-deobfuscator~\cite{javascript-deobfuscator} to deobfuscate JavaScript code. 
We examined the permissions requested in the \texttt{manifest} and verified what (sensitive) information an extension accesses and how it is processed. If any endpoint is contacted to send or retrieve content during execution (e.g., script inclusion, fetch, or XHR), we inspected the payloads of the requests and responses, and we verified with VirusTotal if the domain was malicious. %\footnote{
For complex or large extensions, code analysis was challenging. Therefore, we also examined the runtime behavior of those extensions by installing them in an isolated environment and interacting with them, also using extensions' developer tools for breakpoint debugging and network request monitoring.

\pseudoparagraph{{Manual Effort (qualitative analysis)}} We observed a considerable disparity in the {degree of (apparent) effort attackers invested in crafting their malicious extensions.} As a result, the manual analysis time varied significantly: some extensions required only a few minutes {to make an educated guess}, as merely inspecting the code was sufficient to identify the malicious operations, while others necessitated several hours of detailed examination and demanded the use of multiple tools to uncover the malicious activities, such as debugger or network inspectors.

\begin{table}[]
    \vspace{0.8mm}
    \footnotesize
    \begin{center}
        \resizebox{0.7\columnwidth}{!}{%
            \begin{tabular}{p{7cm}|r r}
                \toprule
                & \textbf{\#Extensions} & \textbf{\#Clusters} \\ \hline
                Proxying network requests through endpoints & 2 & 1 \\
                \midrule
                Navigating the user to {sites VirusTotal reports as malicious} & 3 & 3 \\ \midrule
                Retrieving the list of friends and account details and publishing a Facebook post that includes the CPU model, CPU, and RAM usage automatically and  without permission & 2 & 1 \\
                \midrule
                Requests \& exfiltrate credentials for popular websites & 6 & 1 \\ \midrule
                Exfiltrating the IP address, operating system, and a generated UUID to various endpoints & 22 & 19 \\
                \midrule
                Collecting system information and exfiltrating it to endpoints reported as malicious {on} VirusTotal & 2 & 1 \\
                \midrule
                Ad Blockers that send every {visited URLs} to an endpoint reported as malicious on VirusTotal & 3 & 3 \\
                \midrule
                \textbf{Total} & 40 & 29 \\
                \bottomrule
            \end{tabular}%
        }
    \end{center}
    \caption{Summary of malicious behaviors manually identified}
    \label{tab:malicious_behaviors}
\end{table}

\pseudoparagraph{Results}
Between our data collection (Nov. 2023) and manual analyses (May 2024), 28 of the 1,131 extensions flagged as malicious by our ensemble had already been removed from the CWS by Google for being malicious (overall, these 28 extensions affected $>$2M users before being taken down). Since, these extensions had been confirmed malicious by Google, we did not manually analyze them.
Of the {200} extensions we manually reviewed, we identified \tk{40} malicious extensions that {\small \textit{(i)}}~had not been detected previously and {\small \textit{(ii)}}~were still in the CWS {before our disclosure to Google in May 2024}; these malicious extensions belong to 29 clusters, as shown in \Cref{tab:malicious_behaviors}.
We observe different classes of malicious behavior: some malicious extensions, e.g., redirect users to malicious websites, steal user sensitive data, spy on users, or track them across sites. As a rule, we consider an extension to be malicious when it puts the security or privacy of its users at risk, in line with~\cite{Hsu2024}.
In total, our classifiers could therefore identify (at least) 68 malicious extensions that were unknown to be malicious in Nov. 2023 when we collected our snapshot from the CWS; overall, the 40 malicious extensions still on the CWS in May 2024 affected 11M users (one had $>$5M users).
Besides these, we found an extra 12 extensions with a ``suspicious behavior''. Similarly to~\cite{Kapravelos2014}, suspicious means that we could identify potentially harmful actions and risks to users but without certainty that these represent malicious actions. We have also inspected the Risk Likelihood score provided by Chrome-Stats for our 40 manually-verified malicious extensions: only 11 have a maximum Risk Likelihood of 4, revealing that harmful extensions do ``bypass'' such a metric---and justifying our decision not to rely on it for our~analyses.\footnote{We also manually reviewed 40 extensions with Risk Likelihood=4: at least 5 of these extensions were benign (i.e., FPR$>$12.5\%).}

\pseudoparagraph{Ethical {Disclosure to Google} (and post-acceptance update)}
In May 2024, we disclosed to Google the \tk{40} malicious extensions we identified in the CWS.
{Google responded in August 2024, acknowledging some of our findings. We did not receive any follow up from Google since the last message on August 2024.
Moreover, as of September 2025 (i.e., \textit{after our disclosure}) 17 of our 40 reported extensions have been taken down from the CWS (altogether, these 17 extensions had 1.76M users at the time of their removal) and another 17 have been ``updated'' (altogether, these 17 extensions now have 7.1M users). The remaining 6 extensions that we reported but have not been updated or taken down count 93k users. Hence, and under the assumption that the 17 ``updated'' extensions are not malicious anymore, it can be said that 99\% of the users that had at least one of our reported malicious extensions installed are not affected by such a problem any longer.}

\vspace{1mm}

\summary{Takeaway.}{{Our manual analysis revealed that our ensemble of detectors detected \textit{at least} \tk{68} malicious extensions from the CWS that were unknown to be dangerous when we collected our data in Nov. 2023.} Of those, Google took down 28 extensions (for being malicious) prior to our analyses; another 17 extensions were removed after we disclosed our findings to Google, and 17 have been updated. As of September 2025 only 6 are, unchanged, still on the CWS.} 
\section{Concept Drift \& Browser Extensions}
\label{sec:concept}

\noindent
Our open-world evaluation indicated that our detectors (even if organized in an ensemble) struggle to detect malicious extensions. Even though our manual investigation revealed that our classifiers identified some malicious extensions, the open-world results (in \Cref{subsec:classification_unlabeled}) denoted a stark contrast with the ``near-perfect'' performance achieved during the development phase of our detectors (in \Cref{subsec:classification_results}). We assert that such difference has its root on the intrinsic evolution of the browser extension ecosystem: every day, both benign and malicious extensions ``change'', thereby making it difficult for supervised ML-based classifiers to maintain their performance over long periods of time~\cite{gama2014survey}---a phenomenon typically known as \textit{concept drift}~\cite{barbero2022transcending,apruzzese2022sok}. To the best of our knowledge, no prior work attempted to ``quantify'' the presence of concept drift in the malicious browser extension context (``concept drift'' was only hinted by~\cite{Jagpal2015,kurt2015} in 2015, but never explicitly investigated).

In what follows, we provide factual evidence that concept drift does indeed affect this domain. To this end, we carry out new experiments by proceeding backwards.

\subsection{Preliminary Assessment on \setb}
\label{subsec:concept_unlabeled}

\noindent
As a first step, we examine our last experiments in \Cref{sec:open}. Recall (see \Cref{subsec:data-collection}) that the extensions in \setb{} have been published (or last updated) between Jan.--Nov., 2023. To get a rough idea of the impact of concept drift (which ultimately leads to misclassifications), {we need a more recent dataset of confirmed malicious extensions. As a best-effort approach,} we queried Chrome-Stats in Jan. 2024\footnote{We did this experiment \textit{before} our manual analysis (\Cref{subsection:manual-analysis}).} looking for extensions that have been removed from the CWS for {being malicious} \textit{after November 2023}. We found that 60 of the extensions returned by our query were included in \setb.

Our \meta{} detects 41 (out of 60) extensions as malicious. This represents a false-negative rate of \textit{at least} 31.66\%. This result is in stark contrast with the false-negative rate of only 4.76\% achieved during the development phase of this classifier (\Cref{subsection:metadata_results}).
We observe a similar trend for the \texttt{source code} and the \texttt{combined classifiers} (which, due to parsing errors of Esprima, could only analyze a subset of 41 of these 60 extensions): both of these classifiers can only detect 34 (out of 41) malicious extensions, i.e., a false-negative rate of \textit{at least} 17.07\% (up from 4.61--6.17\%, see Sections~\ref{subsection:sourcecode_results} and \ref{subsection:combined_results}).

Altogether, these results (based on malicious extensions flagged by Google) suggest that our initial hypothesis was likely correct: malicious extensions frequently mutate, thereby leading to evasion even of well-trained classifiers. However, these findings rely on a small sample and cannot strongly support our hypothesis by themselves.

\subsection{Time-aware Evaluation on \seta}
\label{subsec:concept_labeled}
\noindent
To provide further evidence of the existence and effects of concept drift in the browser extension ecosystem, we did another experiment by changing the training and test sets used for developing (and preliminary testing) our detectors. Recall that we created our train and test sets through an 80:20 split by randomly sampling from the entire \seta{} {(see \Cref{subsubsec:datasetA})}, which led to malicious extensions from 2023 being included both in the train and test sets. Hence, to investigate the temporal shifts in malicious extensions, we remove the extensions published in 2023 from the training set, and put them in the test set. This leads to a new training set with \textit{no malicious extensions} published in 2023 (and 80\% ``benign'' extensions from before 2023); and a new test set with all malicious extensions (taken down from the CWS) from 2023, as well as all malicious and benign extensions of the original test set.

\begin{table}[]
    \footnotesize
    \begin{center}
        \resizebox{0.7\columnwidth}{!}{%
            \begin{tabular}{l l r r  r | r}
                \toprule
                 Classifier & Update date... & Benign & Malicious & Total & FNR \\
                 \midrule
                 \texttt{Metadata} & before 2023 & 51 & 1,318 & 1,369 & 3.73\% \\ 
                 \texttt{classifier} & in 2023  & 170 & 144  & 314& 54.14\%\\ 
                 \midrule
                 \texttt{Source code} & before 2023 & 83 & 1,273 & 1,356 & 6.12\%\\
                 \texttt{classifier} & in 2023  & 216 & 59 & 275 & 78.54\%\\
                 \midrule
                 \texttt{Combined} & before 2023 & 51 & 1,305 & 1,356& 3.76\%\\
                 \texttt{classifier} & in 2023  & 182 & 93 & 275& 66.18\%\\
                \bottomrule
            \end{tabular}
        }
    \end{center}
    \caption{Classification results of \textit{malicious} extensions {published or} last updated before vs.\ in 2023; the classifiers have been retrained without any extensions from 2023 in their training set}
    \label{tab:results_without2023}
\end{table}

We retrain our classifiers on the new training set, and then test their performance again. \Cref{tab:results_without2023} shows the predictions of our classifiers on \textit{malicious} extensions from the new test set, differentiating between the publication date of the extensions and the classification output (benign or malicious).
We make two observations.
First, the detection results of malicious extensions last updated \textit{before} 2023 are very similar to those of \Cref{subsec:classification_results}: we have a false-negative rate (FNR) of 3.73--6.12\% compared to 4.61--6.17\% in \Cref{subsec:classification_results}.
Second, the majority of malicious extensions that have been updated \textit{in 2023} are not flagged as malicious by the classifiers: we observe false-negative rates ranging from 54.14\% (\meta) to 78.54\% (\code).
Another intriguing finding is that the \meta{} has a remarkably lower FNR w.r.t. the \texttt{source code} and \comb{}. This may be a sign that attackers are obfuscating their code or at least hiding the harmful functionality of their malicious extensions, while some malicious patterns are still apparent in the metadata.

This additional experiment reinforces our hypothesis of the existence of concept drift in browser extensions. However, insofar, we have only considered \textit{malicious} extensions (and only published in 2023). The following section will provide the last piece of the puzzle, {revealing} that the entire extension ecosystem is (and has been for years) affected by concept drift {(from an ML perspective)}.

\subsection{Further Validation: Longitudinal Analysis}
\label{subsec:concept_validation}

\noindent
``\textit{Is the browser extension ecosystem affected by concept drift?}'' 
To provide {a convincing} answer to this question, we use \seta{} to carry out a longitudinal analysis wherein we train and test the performance of our detectors over time. Inspired by~\cite{pendlebury2019tesseract}, we proceed as follows: given any year $Y$ between 2019--2022, we train our detectors on all extensions {published or} updated before $Y$, and test them on the extensions {published or} updated in $Y$. In this way, we can {approximate a realistic scenario} wherein we progressively assess the performance of our detectors if deployed in the real world {(to analyze ``new'' extensions)} over our considered 4-year timespan---thereby elucidating the degradation (if any) caused by concept drift. We provide in \Cref{fig:distribution} (in the Appendix) the temporal distribution of our extensions in \seta. 
We note that such an experiment is \textit{fair} from an ML perspective: any given extension will only be included either in the train or in the test set---never in both (otherwise, this would naturally inflate the results due to data leakage). {Our assumption is that, if there is no concept drift, then our results should be ``good''.}

\pseudoparagraph{Results}
First, we mention that we follow the exact hyperparameter-optimization procedure discussed in \Cref{subsec:detector_def} to develop our classifiers, and that our cross-validation results (on each training set, and for each classifier) always yielded near-perfect detection performance. However, the outcome of our longitudinal analyses reveals a different story.
We visualize these results in \Cref{fig:conceptdrift-comb} for the \comb{}, and in \Cref{fig:conceptdrift-meta} (in the Appendix) for the \meta. These figures show the TPR, TNR, accuracy (Acc) and precision (1-FDR) across the years. For the \comb{}, we see that the performance is somehow stable in 2019 and 2020 (despite it being retrained at the end of 2019 with the extensions which appeared in 2019).  
However, the performance of \comb{} degrades substantially in the last two years. 
The TPR drops substantially ($\sim$0.5) denoting that malicious extensions present many differences from those ``seen'' during the training phase of our detectors. Moreover, we also note a remarkable drop in the {precision}, with drops to $\sim$0.4 in 2021 and to $\sim$0.3 in 2022. This denotes that more than half of the extensions flagged as malicious by our detectors were actually benign. Such results (which are reflected also by the \meta{}, as shown in Figure~\ref{fig:conceptdrift-meta}) are in stark contrast with those achieved during the {cross-validation} phase, and significantly different from those we presented in \Cref{subsec:classification_results} (see \Cref{fig:perf-clf}). In summary, these ``negative results'' are a strong indicator of concept drift.

\begin{figure}
    \centering
    \includegraphics[width=.95\columnwidth]{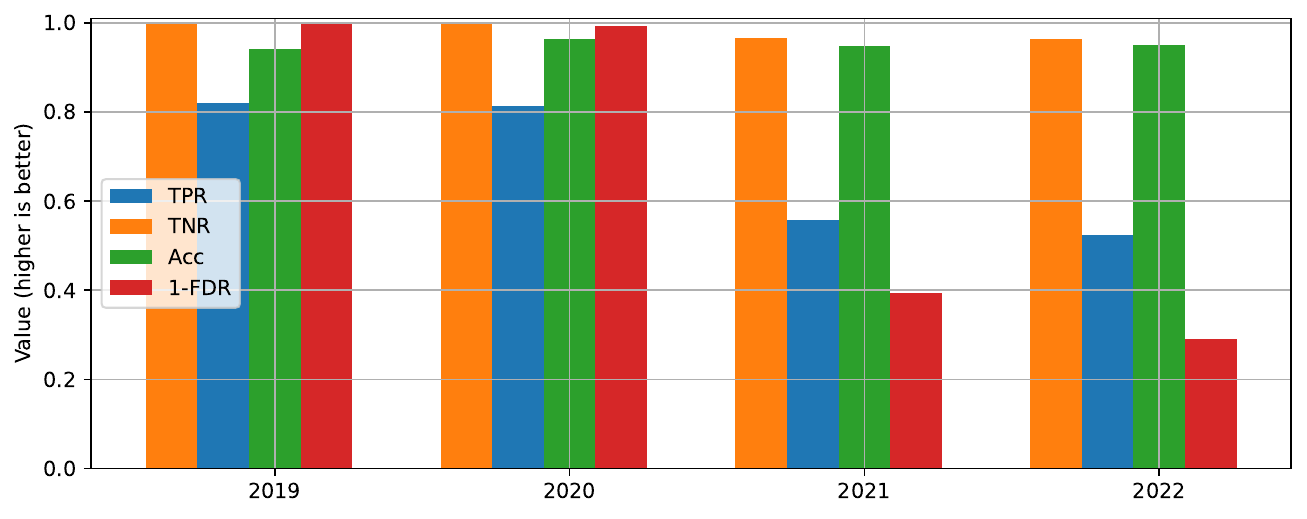}
    \vspace{-4mm}
    \caption{Longitudinal test. We test the \smallcomb{} on the extensions updated every year from 2019--2022 after training it on the extensions {published or} updated in the previous years} 
    \label{fig:conceptdrift-comb}
    \vspace{-2mm}
\end{figure}

\begin{table*}[t]
    \centering
    \resizebox{\columnwidth}{!}{
        \begin{tabular}{c?l|c?l|c?l|c?l|c?l|c}
            \bottomrule
            \multirow{2.5}{*}{\begin{tabular}{c}
                 \textbf{Rank} \\ (top-5)
            \end{tabular}} & \multicolumn{2}{c?}{2019} & \multicolumn{2}{c?}{2020} & \multicolumn{2}{c?}{2021} & \multicolumn{2}{c?}{2022} & \multicolumn{2}{c}{2023} \\
            \cmidrule{2-11}
            & 
            \textbf{Feature Name} & \textbf{Value} & 
            \textbf{Feature Name} & \textbf{Value} & 
            \textbf{Feature Name} & \textbf{Value} & 
            \textbf{Feature Name} & \textbf{Value} & 
            \textbf{Feature Name} & \textbf{Value} \\
            \toprule

            \#1 & 
            {\footnotesize fileCount} & {\footnotesize 0.0272} & 
            {\footnotesize fileCount} & {\footnotesize 0.0271} & 
            {\footnotesize fileCount} & {\footnotesize 0.0241} &
            {\footnotesize fileCount} & {\footnotesize 0.0213} &
            {\footnotesize fileCount} & {\footnotesize 0.0206} \\

            \#2 & 
            {\footnotesize Full-Summary LOVE} & {\footnotesize 0.0212} & 
            {\footnotesize Permissions topSites} & {\footnotesize 0.0245} & 
            {\footnotesize Full-Summary WALLPAPER} & {\footnotesize 0.0223} &
            {\footnotesize Full-Summary WALLPAPER} & {\footnotesize 0.0196} &
            {\footnotesize Full-Summary FAVORITE} & {\footnotesize 0.0154} \\
            
            \#3 & 
            {\footnotesize Full-Summary HIGH} & {\footnotesize 0.0211} & 
            {\footnotesize Full-Summary THEME} & {\footnotesize 0.0196} & 
            {\footnotesize size} & {\footnotesize 0.0178} &
            {\footnotesize size} & {\footnotesize 0.0170} &
            {\footnotesize Description NEW} & {\footnotesize 0.0151} \\
            
            \#4 & 
            {\footnotesize (source code \#157)} & {\footnotesize 0.0209} & 
            {\footnotesize Full-Summary WALLPAPER} & {\footnotesize 0.0185} & 
            {\footnotesize Full-Summary FAVORITE} & {\footnotesize 0.0170} &
            {\footnotesize Description NEW} & {\footnotesize 0.0146} &
            {\footnotesize Full-Summary TIME} & {\footnotesize 0.0131} \\
            
            \#5 & 
            {\footnotesize (source code \#105)} & {\footnotesize 0.0171} & 
            {\footnotesize Full-Summary FAVORITE} & {\footnotesize 0.0175} & 
            {\footnotesize Description TAB} & {\footnotesize 0.0164} &
            {\footnotesize Full-Summary THEME} & {\footnotesize 0.0126} &
            {\footnotesize size} & {\footnotesize 0.0117} \\
            \bottomrule
        \end{tabular}
    }
    \caption{Evolution of the \textit{feature importance} (of \smallcomb) after retraining it yearly on \smallseta{}}
    \label{tab:feature_importance}
\end{table*}

\pseudoparagraph{Feature Importance}
Previously, we discussed \textit{how} browser extensions are affected by concept drift (by looking at the misclassifications of the detectors); now we focus on \textit{why} extensions are affected by concept drift by analyzing the feature importance learned by our classifiers over the years. This is instructive to highlight ``changes'' in the data distribution, which may explain the source of concept drift. Hence, we retrieve the ranking of the top-5 most important features of the \comb{} (which our initial assessment in \Cref{subsec:classification_results} showed to perform best) after every year; we also report it for 2023 (i.e., by training the \comb{} on extensions that appeared before Jan. 1, 2023). The results are in Table~\ref{tab:feature_importance}. We see that the ``fileCount'' feature is always ranked 1st. However, we also observe many changes over the years: the ``Full-Summary WALLPAPER'' feature is ranked 4th in 2020 and 2nd in 2021 and 2022, but is outside the top-5 in 2019 and 2023; ``Permission topSites'' is ranked 2nd in 2020, and outside the top-5 in any other years; source-code features (two) are in the top-5 only in 2019.
This analysis indicates that, over the years, the overall feature distribution  of malicious and benign extensions changes. Such dynamics explain \textit{why} our detectors have an underwhelming performance when tested ``outside the lab'' to analyze extensions in the real world---on the CWS.

\vspace{1mm}

\summary{Takeaway}{We {provided factual evidence} that the browser extension ecosystem is also impacted by concept drift {and discussed reasons why this detrimental phenomenon occurs}. To reduce its impact, detectors need to be retrained on a regular basis---requiring a constant stream of new (and correctly labeled) benign and malicious extensions {(see suggestions in \Cref{subsec:active}).}}
\section{Additional Experiments (and Potential Countermeasures)}
\label{sec:extra}

\noindent
Our datasets and classifiers can be used to shed further light on the evolution of the browser-extension landscape from a security viewpoint. 
We take this opportunity to carry out three complementary experiments, summarised below and discussed in the following subsections.
\begin{itemize}[leftmargin=*]
    \item \textbf{Susceptibility of Extensions to Concept Drift.} We investigate if certain \textit{categories} (or \textit{manifest versions}) of extensions lead to more prominent misclassifications, indicating a more impactful drift. For benign (resp. malicious) extensions, the ``search-tools'' (resp. ``photos'') category tends to have a very unstable behavior; moreover, the release of Manifest V3 (MV3) had a substantial impact on the detection of malicious extensions---which is an expected result.
    \item \textbf{Active Learning to Mitigate Concept Drift.} 
    Active learning is a well-known technique to optimize the labeling efforts that can mitigate the impact of concept drift~\cite{apruzzese2022sok}. However, active learning has never been explored in the browser-extension ecosystem. We are the first to do so. We found that the best returns are achieved by re-training the classifiers monthly on 15 labeled extensions (``actively-suggested'' via \textit{uncertainty sampling}). 
    \item \textbf{Alternative Design Choices for our Classifiers.} We explored {\small \textit{(i)}}~if the performance of our classifiers could be improved by better balancing the training sets---but none of our experiments were successful; and also {\small \textit{(ii)}}~if it was possible to minimize the FPR (at the expense of the TPR) by tinkering with the detection threshold of our random-forest classifiers. We found that (in our ``lab setup'') we can reduce the FPR from 0.0107 to 0.00047 by sacrificing only 5\% of the TPR.
\end{itemize}
The code of each experiment is provided in our repository~\cite{repository}.

\subsection{Which Browser Extensions are more Impacted by Concept Drift?}
\label{subsec:impact}

\noindent
Given that we have complete information on each extension, provided by Chrome-Stats, it is instructive to carry out a low-level analysis to measure the impact of concept drift on specific clusters of extensions. Specifically, we focus the attention on two aspects: {\small \textit{(i)}}~the \textit{category} of extensions, and {\small \textit{(ii)}}~the Manifest version.

\pseudoparagraph{Categories of extensions} The goal of this analysis is to investigate the performance over time of our \comb, but on specific categories of extensions: categories for which the performance is underwhelming are those that are most likely affected by concept drift. [\textit{Method}] For clarity, we consider the nine most-prevalent categories of extensions included in \seta{}: ``productivity'', ``fun'', ``web-development'', ``communication'', ``accessibility'', ``shopping'', ``photos'', ``news'', and ``search-tools''; we aggregate all remaining extensions (21 in total, i.e., less than 0.1\% of \seta{}) in a single ``other'' group. We report in Table~\ref{tab:categories_full} the number of extensions for each group (and for each year), as well as the accuracy achieved by the \comb{} on the respective category of extensions.\footnote{For these experiments, we use the same setup as in Section~\ref{subsec:concept_validation} (we simply breakdown the results according to the specific category). However, we use a fixed threshold=8.8\% because the one of 9.2\% was derived by analysing data until 2023, whereas here the training data changes incrementally. For simplicity and consistency, we hence use a slightly different threshold.}
[\textit{Analysis}] We see that, in the case of benign extensions, the three most problematic categories are ``photos'', ``fun'', and ``communications'': apparently, benign extensions of these categories become more similar to previously seen \textit{malicious} extensions. In contrast, for malicious extensions, the most drifting categories are ``search-tools'', ``productivity'', and ``communications'': the classification accuracy is $\approx$50\% every year. Nonetheless, and also as evidenced by the recent extensive analysis by Hsu et al.~\cite{Hsu2024}, certain categories of extensions do not have many malicious extensions (e.g., for ``news'', only 6 extensions are malicious).

\begin{table}[h]
    \centering
    \renewcommand{\arraystretch}{1.2}
    \resizebox{\textwidth}{!}{
    \begin{tabular}{ll|rr|rr|rr|rr}
        \toprule
        \multirow{3}{*}{\textbf{Category}} & \multirow{3}{*}{\textbf{Class}} & \multicolumn{2}{c|}{\textbf{2019}} & \multicolumn{2}{c|}{\textbf{2020}} & \multicolumn{2}{c|}{\textbf{2021}} & \multicolumn{2}{c}{\textbf{2022}} \\
        \cline{3-10}
        
        & & Accuracy & Total & Accuracy & Total & Accuracy & Total & Accuracy & Total \\
        \midrule
        productivity & Benign & \cellcolor{green!20} 100.00\% & 2941 & \cellcolor{green!20} 99.86\% & 4417 & \cellcolor{green!20} 97.51\% & 5784 & \cellcolor{green!20} 97.33\% & 7308 \\
        & Malicious & 0.00\% & 34 & 27.38\% & 168 & 39.25\% & 186 & 47.86\% & 117 \\
        fun & Benign & \cellcolor{green!20} 100.00\% & 899 & \cellcolor{green!20} 99.60\% & 1011 & \cellcolor{green!20} 90.53\% & 1151 & \cellcolor{green!20} 89.35\% & 1605 \\
        & Malicious & 1.08\% & 9 & 30.65\% & 124 & \cellcolor{green!20} 77.17\% & 127 & \cellcolor{green!20} 83.33\% & 90 \\
        web-development & Benign & \cellcolor{green!20} 100.00\% & 886 & \cellcolor{green!20} 99.90\% & 1048 & \cellcolor{green!20} 98.36\% & 1160 & \cellcolor{green!20} 98.79\% & 1488 \\
        & Malicious & 0.00\% & 30 & 12.50\% & 16 & 26.67\% & 15 & 46.67\% & 15 \\
        communication & Benign & \cellcolor{green!20} 100.00\% & 727 & \cellcolor{green!20} 99.79\% & 969 & \cellcolor{green!20} 96.20\% & 1027 & \cellcolor{green!20} 95.87\% & 1138 \\
        & Malicious & 0.00\% & 18 & 13.11\% & 61 & 47.83\% & 23 & 29.63\% & 81 \\
        accessibility & Benign & \cellcolor{green!20} 99.85\% & 651 & \cellcolor{green!20} 100.00\% & 870 & \cellcolor{green!20} 96.74\% & 1195 & \cellcolor{green!20} 97.65\% & 1447 \\
        & Malicious & 0.00\% & 11 & 68.97\% & 261 & 50.00\% & 44 & 34.04\% & 47 \\
        shopping & Benign & \cellcolor{green!20} 100.00\% & 335 & \cellcolor{green!20} 100.00\% & 541 & \cellcolor{green!20} 98.44\% & 771 & \cellcolor{green!20} 98.17\% & 763 \\
        & Malicious & - & 0 & - & 0 & \cellcolor{green!20} 93.10\% & 13 & 14.29\% & 7 \\
        photos & Benign & \cellcolor{green!20} 98.04\% & 51 & \cellcolor{green!20} 100.00\% & 70 & \cellcolor{green!20} 72.48\% & 109 & \cellcolor{green!20} 77.86\% & 98 \\
        & Malicious & \cellcolor{green!20} 82.95\% & 2105 & \cellcolor{green!20} 99.38\% & 70 & \cellcolor{green!20} 88.89\% & 9 & \cellcolor{green!20} 100.00\% & 8 \\
        news & Benign & \cellcolor{green!20} 100.00\% & 102 & \cellcolor{green!20} 100.00\% & 164 & \cellcolor{green!20} 96.89\% & 193 & \cellcolor{green!20} 98.16\% & 163 \\
        & Malicious & \cellcolor{green!20} 100.00\% & 2 & 0.00\% & 2 & \cellcolor{green!20} 100.00\% & 2 & \cellcolor{green!20} 100.00\% & 2 \\
        search-tools & Benign & \cellcolor{green!20} 100.00\% & 1 & \cellcolor{green!20} 100.00\% & 3 & \cellcolor{green!20} 100.00\% & 1 & \cellcolor{green!20} 100.00\% & 2 \\
        & Malicious & \cellcolor{green!20} 99.87\% & 763 & 37.50\% & 40 & 50.00\% & 28 & 63.89\% & 36 \\
        (others) & Benign & \cellcolor{green!20} 100.00\% & 11 & \cellcolor{green!20} 100.00\% & 1 & \cellcolor{green!20} 100.00\% & 1 & \cellcolor{green!20} 100.00\% & 2 \\
        & Malicious & - & 0 & - & 0 & - & 0 & \cellcolor{green!20} 100.00\% & 1 \\
        \midrule
        \textbf{TOTAL} & Benign & \cellcolor{green!20} 99.97\% & 6514 & \cellcolor{green!20} 99.85\% & 9093 & \cellcolor{green!20} 96.51\% & 11391 & \cellcolor{green!20} 96.35\% & 14056 \\
        & Malicious & \cellcolor{green!20} 82.04\% & 3061 & \cellcolor{green!20} 81.25\% & 2139 & 55.70\% & 1465 & 52.49\% & 402 \\
        \bottomrule
    \end{tabular}
    }
    \caption{Accuracy (by \comb{}) and number of extensions for every year (cells in green highlight an accuracy over 70\%)}
    \label{tab:categories_full}
    \vspace{-2mm}
\end{table}

\begin{table}[t]
    \centering
    \resizebox{0.7\textwidth}{!}{
    \renewcommand{\arraystretch}{1.2}
    \begin{tabular}{ll|rr|rr|rr|rr}
        \toprule
        \multirow{3}{*}{\textbf{Manifest}} & \multirow{3}{*}{\textbf{Class}} & \multicolumn{2}{c|}{\textbf{2019}} & \multicolumn{2}{c|}{\textbf{2020}} & \multicolumn{2}{c|}{\textbf{2021}} & \multicolumn{2}{c}{\textbf{2022}} \\
        \cline{3-10}
        & & \multicolumn{2}{c|}{Benign} & \multicolumn{2}{c|}{Malicious} & \multicolumn{2}{c|}{Benign} & \multicolumn{2}{c}{Malicious} \\
        \cline{3-10}
        & & Accuracy & Total & Accuracy & Total & Accuracy & Total & Accuracy & Total \\
        \midrule
        MV2 & Benign & 99.97\% & 6,514 & 99.86\% & 9,093 & 95.93\% & 9,377 & 94.27\% & 3,841 \\
        & Malicious & 82.03\% & 3,061 & 81.25\% & 2,139 & 61.43\% & 407 & 83.13\% & 83 \\
        MV3 & Benign & - & - & - & - & 99.21\% & 2,014 & 97.13\% & 10,215 \\
        & Malicious & - & - & - & - & 15.52\% & 58 & 44.51\% & 319 \\
        \bottomrule
    \end{tabular}
    }
    \caption{Accuracy (as by the \comb{}) and total number of extensions per manifest version (MV2 vs.\ MV3) from 2019 to 2022}
    \label{tab:manifest}
    \vspace{-4mm}
\end{table}

\pseudoparagraph{Manifest version}
We now focus on the current Manifest version (MV3), whose introduction in 2021 led to a substantial change in the browser extension ecosystem. Our goal is quantifying the extent to which the release of MV3 impacted the performance of our \comb{}. We report the results of this study in Table~\ref{tab:manifest}. We find it intriguing that the performance of our \comb{} on \textit{benign} extensions has been minimally impacted by the release of MV3 given that the accuracy is always above 97\%. However, for malicious extensions, the situation is substantially different: while the \comb{} could detect over 80\% of malicious extensions in both 2019 and 2020 (all being MV2 extensions), the detection accuracy on malicious MV3 extensions dropped to 15\% in 2021 and only got slightly better (to 44\%) in 2022. Such a fluctuation is because the \comb{} was not trained on any MV3 extensions when they came out in 2021, which explains the underwhelming performance; after training the \comb{} on the (few) malicious MV3 extensions that appeared in 2021, its performance improved---but over 55\% of malicious MV3 extensions released in 2022 still bypass its detection. Put simply, this finding confirms the existence of concept drift in the browser extension ecosystem.

\subsection{What are some Ways to Mitigate the Effects of Concept Drift?}
\label{subsec:active}

\noindent
It is evident that the browser-extension ecosystem is affected by a concept-drift problem: the constant changes in new browser extensions make it hard for classifiers based on supervised ML to retain practical performance. However, a promising strategy to mitigate the disruptive effects of such changes (which affect both benign and malicious extensions) is that of \textit{active learning}~\cite{braun2024understanding}. Hence, we will now assess how practical such techniques are in this domain via original experiments. We stress that, to the best of our knowledge, we are the first to empirically evaluate active-learning methods in the context of malicious browser extension detection---hence, our approach can be considered as an original technical contribution.

\pseudoparagraph{Design Choices} First, we consider the application of active learning based on \textit{uncertainty sampling}~\cite{apruzzese2022sok, pendlebury2019tesseract}. At a high level, active learning seeks to optimize the development process of ML-based classifiers by ``actively suggesting'' a small number of datapoints (in our case, extensions) that should be labeled---and then update the classifier by re-training it on these (new) labeled samples. The idea is that some extensions are more informative than others for detection purposes, and hence having the analyst label such ``very informative extensions'' can result in a better learning phase of the corresponding classifier. In the case of uncertainty sampling, the suggestions are provided by considering extensions for which the classifier is ``uncertain'': indeed, extensions for which the detector is certain that they are malicious (or benign) are likely very similar to those in the training set; in contrast, extensions for which the detector is uncertain are likely to present characteristics that would help the classifier to refine its decision boundaries.

\pseudoparagraph{Implementation} We carry out our experiments as follows. First, we consider a time-period of 12 months and, specifically, the entire 2022 year. This is because it is the most recent year for which we have complete availability of extensions with ground truth (see Section~\ref{sec:data}). Then, we consider only the \comb{} (with the fixed threshold used in §\ref{subsec:impact}), since it is the classifier with the best performance (according to Section~\ref{subsection:combined_results}). In this setup, we then iteratively test and update (by means of active learning) the detector after every month. Specifically:
\begin{itemize}[leftmargin=*]
    \item we begin by training the \comb{} by considering all extensions published (or last updated) before January 1st, 2022 (54,056 extensions);
    \item then, we test the detector on the extensions published (or last updated) in January 2022. Afterwards, we take the ``probabilities'' (provided by the \textit{predict\_proba} method of scikit-learn) and rank all these extensions according to those that are closer to the decision boundary---i.e., representing those extensions for which the classifier is the most uncertain;
    \item we select the $K$ top-ranked extensions, and put them in the training set (by assigning the correct label). Finally, we retrain the ``active-learning updated''  \comb{} (having 54,056+$K$ extensions) and test it on extensions published (or last updated) in February 2022.
\end{itemize}  
We repeat the procedure above for every month of 2022, i.e., in December 2022 our detector will be trained on a dataset of 56,056+(11*$K$) extensions---so that, every month, only the best $K$ extensions are used to update the \comb{}. 

\begin{figure*}[!t]
\centering
\begin{subfigure}{0.31\textwidth}
\centering
\includegraphics[width=\columnwidth]{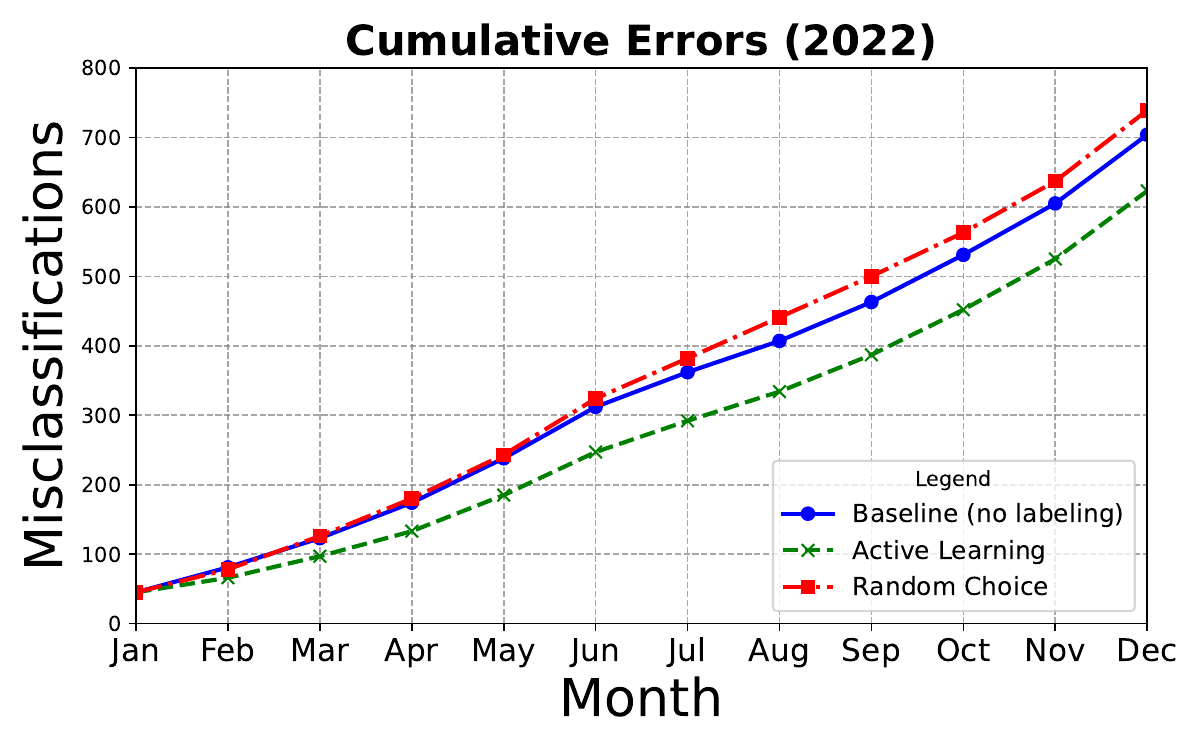}

\caption{Cum. Errors (lower is better)}
\label{sfig:al_errors}
\end{subfigure}
\begin{subfigure}{0.31\textwidth}
\centering
\includegraphics[width=\columnwidth]{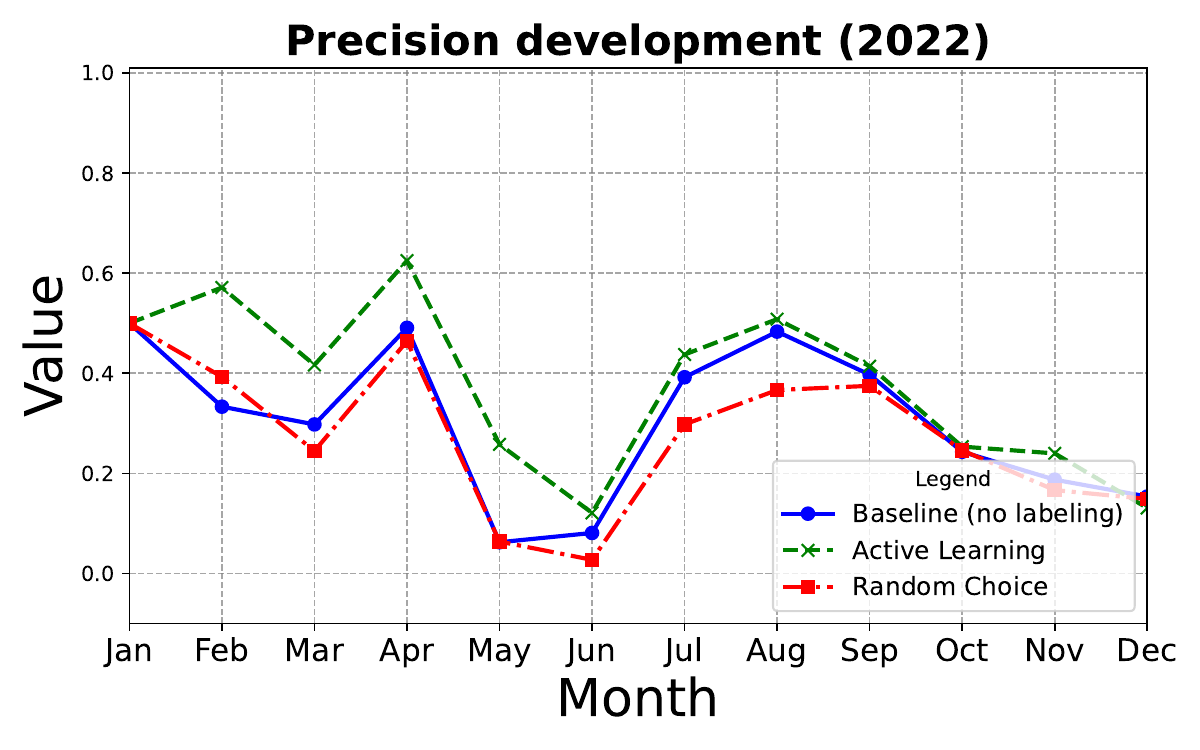}

\caption{Precision (higher is better)}
\label{sfig:al_precision}
\end{subfigure}
\begin{subfigure}{0.31\textwidth}
\centering
\includegraphics[width=\columnwidth]{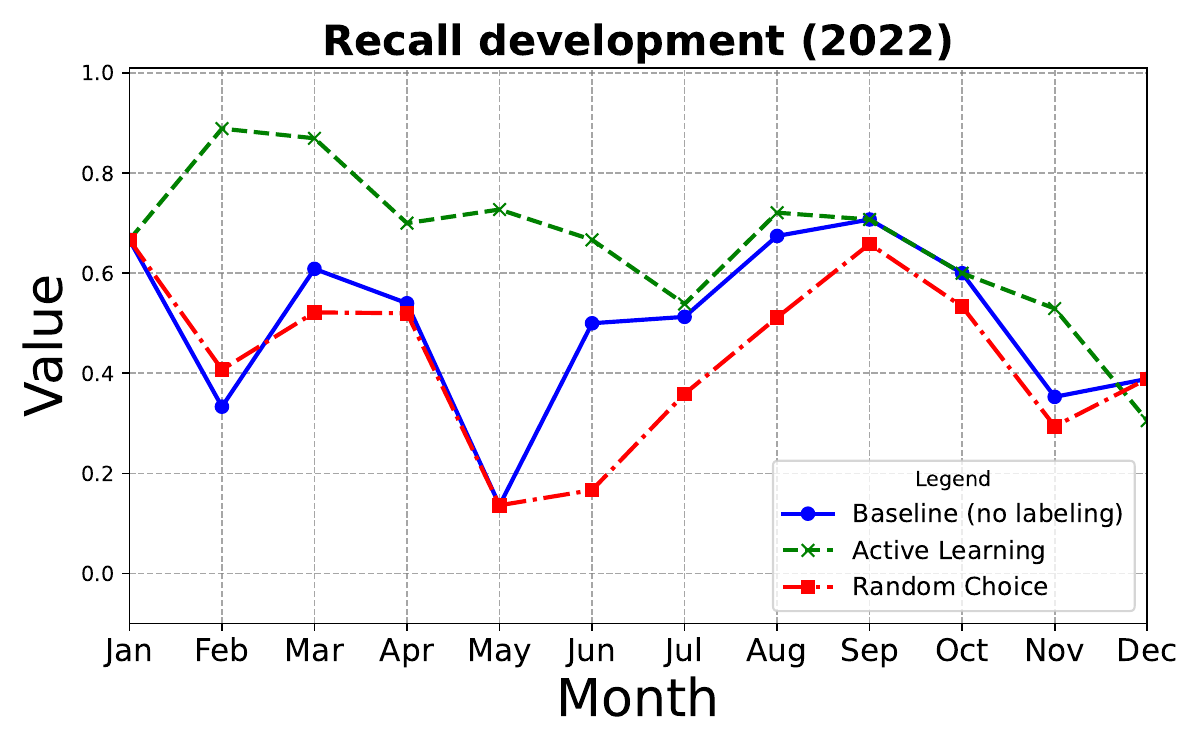}

\caption{Recall (higher is better)}
\label{sfig:al_recall}
\end{subfigure}
\vspace{-4mm}
\caption{Experiments with active learning. The blue line corresponds to the \smallcomb{} trained on all extensions before Jan 1st 2022 (and never re-trained).
The green line corresponds to the \smallcomb{} that is re-trained monthly on an extra 15 extensions for which the model was \textit{``the most uncertain''} the previous month.
The red line corresponds to the \smallcomb{} that is monthly re-trained by including an extra 15 \textit{randomly chosen} extensions of the previous month.
}
\label{fig:activelearning}
\vspace{-2mm}
\end{figure*}

\pseudoparagraph{Evaluation and Results} For comparison, we consider the performance of the ``baseline'', representing the \comb{} trained only with the extensions before January 1st, 2022; and also consider the performance of the \comb{} trained by randomly choosing $K$ extensions (instead of using active learning) every month, which can be considered an ablation study to ascertain the benefit of our active-learning implementation~\cite{apruzzese2022sok}. The performance metrics of choice are the precision and recall (measured on a monthly basis) as well as the cumulative errors (i.e., the overall number of misclassifications across the entire test window). We tested our implementation by considering various values of $K$: 10, 15, 25, 50; however, we obtained the best results (in terms of labeling effort and performance gains) with $K$=15 (for instance, we found no substantial improvement for $K$=50 despite requiring more than 3x the labeling effort). We report the results for $K$=15 in Figure~\ref{fig:activelearning}. We can see that overall, \textit{active learning provides a substantial benefit compared to the baseline} (and also compared to the random choice), which is evident by observing the cumulative errors (\Cref{sfig:al_errors}). However, in some cases (e.g., for June 2022, see the drop in \Cref{sfig:al_precision}) the ``new'' extensions are so different from those which appeared before that our detector struggles even after the update.
Another intriguing finding is that \textit{updating the classifier by randomly drawing extensions can be detrimental}: in most plots, the ``random choice'' (red line) is worse than the ``baseline'' (blue line), despite the fact that the ``random choice'' is trained on a superset of the training set of the ``baseline''.
This is why we recommend developers to adopt active-learning practices if they seek to deploy (and update) automatic classifiers based on supervised ML.

\subsection{Alternative Design Choices for our Classifiers}
\label{subsec:falsepositive}

\noindent
To develop the classifiers used in our analysis, we adopted various design choices. Here, we describe some experiments wherein we explore alternative ways to develop our classifiers---which can be instructive for future work that seeks to improve our detectors.

\pseudoparagraph{Sub-sampling}
The number of benign extensions in \seta{} is over 9 times larger than that of malicious extensions. To account for such class-imbalance, we set a ``balanced'' class weight when we defined our classifiers in scikit-learn. However, we have also experimented by considering subsets of (randomly drawn) benign extensions. Specifically, from our training set (i.e., 80\% of \seta{}), we isolate 10\%, 20\%, 40\%, 80\% of the available benign extensions, which are used alongside all the available malicious extensions to train the \comb{}. We then measure the performance on the test set (i.e., 20\% holdout of \seta{}) and repeat the experiment five times to account for potential bias introduced by the random draw. We achieved the following results:
\begin{itemize}[leftmargin=*]
    \item For 10\% (i.e., an equal composition of benign\,/\,malicious): \textit{Acc}=0.60$\pm$0.01; \textit{FPR}=0.43$\pm$0.01, \textit{1-FDR}=0.20$\pm$0.01, \textit{TPR}=0.99$\pm$0.005;
    \item For 20\% (i.e., benign are twice the malicious): \textit{Acc}=0.81$\pm$0.005; \textit{FPR}=0.20$\pm$0.005, \textit{1-FDR}=0.34$\pm$0.005, \textit{TPR}=0.98$\pm$0.001;
    \item For 40\% (i.e., benign are four times the malicious): \textit{Acc}=0.93$\pm$0.01; \textit{FPR}=0.07$\pm$0.01, \textit{1-FDR}=0.62$\pm$0.01, \textit{TPR}=0.97$\pm$0.01;
    \item For 80\% (i.e., benign are eight times the malicious): \textit{Acc}=0.98$\pm$0.01; \textit{FPR}=0.01$\pm$0.01, \textit{1-FDR}=0.86$\pm$0.01, \textit{TPR}=0.96$\pm$0.01.
\end{itemize}
In light of these results, we can hence conclude that our choice of using all benign samples, and accounting for imbalance via the ``balanced'' class weight parameter, was a valid choice. We recommend not to adopt subsample strategies: doing so may lead to classifiers which are not provided with vital information to detect malicious extensions.

\pseudoparagraph{Minimizing the FPR}
Our classifiers minimize the cumulative sum of false positives and false negatives. However, it is legitimate to opt for alternative strategies focused on, e.g., minimizing the rate of false positives. We explored such a possibility by choosing a different threshold and assessing its effects. We did this for the \comb{}, given that it was the best detector. By looking at the plots in Figure~\ref{fig:threshold}, we chose a higher threshold=35\% (instead of 9.2\%, used in the experiments done in this paper), and we tested the resulting performance on the same test set (i.e., 20\% of \seta{}); we also repeated this experiment five times to account for any potential fluctuation. We found that such a threshold leads to a detector with TPR=0.9054 and FPR=0.00047 (compared to a TPR=0.9551 and FPR=0.0096 achieved with a threshold of 9.2\%). In other words, it is possible to substantially reduce the FPR at the cost of only 5\% in the TPR. Therefore, future work that seeks to develop detectors that minimize the FPR can use such a strategy to better comply with certain requirements.

\section{Discussion and Recommendations}
\label{sec:discussion}

\noindent
We first summarize our major findings and then discuss the limitations of our research. 
Finally, we make some considerations on the robustness of our detectors to targeted evasion attempts.

\subsection{Major Findings}
\label{subsec:findings}

\noindent
We elucidated three novel and complementary discoveries.

First, in \Cref{sec:labeled-m-ext-detect}, we developed three classifiers (using a mix of novel feature engineering and prior work) to detect malicious browser extensions. We showed that these classifiers performed well on a dataset of real browser extensions collected from the official gallery (see~\Cref{sec:data}), and our extracted features also provide insights on underlying properties of benign/malicious extensions. Moreover, we also showed that our detectors can be leveraged to identify malicious extensions ``overlooked'' by Google on the CWS (we identified 68 malicious extensions, impacting over 13M users). We publicly release the implementation of our classifiers~\cite{repository}. This contribution is an important stepping stone for future work: we are not aware of open-source detectors of malicious browser extensions that have been validated on large datasets of browser extensions (we discuss related work in \Cref{sec:related}).

Second, in \Cref{sec:open}, we found that automated solutions for analyzing browser extensions---with the aim of detecting \textit{malicious} ones---present limitations. This spans not only our custom-developed classifiers, but also supervised-ML-based detectors proposed by prior work (e.g., \jast{}~\cite{JaSt} ultimately seeks to detect malicious JavaScript, which is a crucial component in browser extensions) as well as commercial tools such as VirusTotal and CRXcavator. We also observe that even though VirusTotal had been shown to be not very reliable by DeKoven et al.~\cite{DeKoven2017}, their analysis was carried out in 2017, i.e., over 8 years ago; in contrast, our analysis is more recent. Altogether, these findings underscore that \textit{automatically} detecting malicious browser extensions is a fundamentally hard problem and that manual analysis is still needed for accurate verification of ground truth.

Third, the collective findings in \Cref{sec:concept} (on both \seta{} and \setb{}), reveal that concept drift significantly affects the browser extension ecosystem---explaining why applying supervised-ML for automated detection of malicious browser extensions is challenging.\footnote{Note that the decreased performance is not due to ``overfitting''~\cite{dietterich1995overfitting}. Our experiments in our lab setting showed good performance on a test set disjoint from the training set---which is the correct evaluation protocol by assuming that the i.i.d. assumption holds. Our time-aware analysis revealed that the issue stems from the fact that the generative process of {\texttt{\scriptsize Dataset L}} does not follow an i.i.d. assumption, but rather follows the chaotic evolution of the browser-extension ecosystem. Hence the samples in the test set should be drawn from those that chronologically follow the samples in the training set.} Over time, malicious extensions become ``more similar'' to benign extensions that appeared previously; and, in turn, some benign extensions also become ``more similar'' to malicious ones---leading to misclassifications that degrade the overall performance of detectors based on supervised ML.\footnote{Note: the purpose of our research was to examine the feasibility of applying supervised ML to detect malicious browser extensions. Our findings indicate that doing so is not easy. We acknowledge that there exist other ML-based approaches that could potentially be used for the same purpose (e.g., some works proposed using LLMs in the browser-extension context, but for ancillary tasks~\cite{nayak2024experimental}). Yet, our results suggest that any sort of ``machine-based'' detector requires constant updating due to the dynamic nature of the browser-extension ecosystem.} This discovery, which echoes the results achieved by prior work in other cyber security applications of supervised ML classifiers (e.g., Android or Windows malware~\cite{jordaney2017transcend}, intrusion detection~\cite{yang2021cade}), emphasizes the need for more work in the browser-extension context to address this problem.

\summary{Remark.}{Our last finding is \textit{fair}. When we begun our research, we did not expect to find such a strong concept-drift effect. The detectors used in our analysis (both the novel ones and those based on prior work) achieved good performance on \seta{}. We thought our initial goal (i.e., developing automated and practical detectors of malicious browser extensions by relying on supervised ML) was achieved---but we were wrong: despite being automated, the performance of our detectors is underwhelming in practice.}

\subsection{Limitations}
\label{subsec:limitations}

\noindent
For our research, we relied on various ``best-effort'' strategies, and we encountered some errors in the implementation of some methods. However, we have reason to believe that none of such occurrences threaten the validity of our conclusions.

From an ML perspective, it is desirable if a classifier is trained on a ``sufficiently large''~\cite{apruzzese2022sok} number of datapoints---and, ideally, if the various classes are relatively balanced. Unfortunately, collecting a meaningful dataset of \textit{verified} benign and malicious extensions \textit{for realistic} experiments is challenging. 
The analysis carried out in our paper relies on a set of 106,200 extensions, of which 7,140 are known to be malicious, while the remainder are ultimately unverified:\footnote{Such a lack of ground truth prevents one from using existing tools/methods to \textit{reliably} find out if an extension is malicious or not (and our assessment on ``commercial'' detectors revealed that existing tools are not very accurate).} we assumed (and justified---see \Cref{subsec:datasets}) that those (63,598) whose last update occurred before Jan. 1, 2023 are ``benign'', but such an hypothesis may not hold universally, and we cannot exclude that our choice may have impacted our results. However, our choice is based on the findings of a very recent work~\cite{Hsu2024} revealing that only dozens of malicious extensions remain in the CWS \textit{for years}. Given that our \seta{} contains 63k extensions (from the CWS) whose last update spans across 2012--2022, the likelihood that the number of mislabeled benign extensions in \seta{} could have impacted our results to the point of threatening our conclusions is negligible. For instance, our temporal analysis does show that malicious extensions (for which we do have verified ground truth) are subject to frequent changes over the years. Nonetheless, it is well known that providing accurate ground truth is hard~\cite{braun2024understanding, apruzzese2022sok}, and even practitioners adopt coarse labeling strategies~\cite{van2022deepcase, braun2024understanding}. Nevertheless, we could have ``artificially'' increased the number of extensions (thereby extending our datasets) by considering individual versions of an extension as a stand-alone extension: we did not do so because such an approach would skew the results, since it would strongly violate the ``independent and identically distributed random variables'' assumption of ML~\cite{apruzzese2022sok}.

As justified in \Cref{subsec:components}, we rely on \texttt{manifest}, content scripts, and service worker (and Chrome-Stats) to extract the features to detect malicious extensions. Our choice is confirmed by our empirical analysis (only 7\% of the malicious extensions in Chrome-Stats are malicious despite not having these files). Our analysis thus provides a lower bound of malicious extensions. In other words, there may be other ways to develop detectors of malicious browser extensions via supervised ML. However, our results (in \Cref{subsec:classification_results}) suggested a good performance of our classifiers, thereby justifying our subsequent analyses. Nonetheless, we leave the investigation of malicious behaviors included in elements not considered in our analysis to future work. Our publicly available tools~\cite{repository} should facilitate development of such ``enhanced'' detectors.

As mentioned in \Cref{subsubsec:source_code_features}, parsing errors of Esprima prevented us from analyzing some extensions with \jast~\cite{JaSt}. We reduced the syntax errors by: {\small \textit{(i)}}~updating the Esprima version used by \jast{} in 2018, which enabled us to analyze new constructs such as Optional Chaining; and {\small \textit{(ii)}}~concatenating content scripts with service worker and putting them in two separate \texttt{BlockStatement}, which avoided errors for duplicate constant values.
Out of 106,200 extensions, Esprima could not parse 7,448. Given our goals, we did not omit these, which could still be analyzed by \meta. Moreover, {as co-authors of \jast{}~\cite{JaSt}, we confirm that using \jast{} to detect malicious source code is sensible in the browser-extension context, also because browser extensions do not even use as many code-transformation techniques as client-side JavaScript}.

\subsection{Robustness Considerations}
\label{subsec:robustness}
\noindent
In our threat model (see Section~\ref{subsec:scope}), we envision an attacker that does not deliberately attempt to evade our detectors. This is motivated by the fact that, to the best of our knowledge, we are the first to develop (and openly share) detectors powered by supervised ML to flag malicious extensions. However, we acknowledge that real-world attackers may resort to adversarial tactics to bypass our detectors---especially in a white-box setting (which is made possible given our public release of our implementation). Here, we discuss some ways attackers may use to have their malicious extensions be classified as benign by our detectors. We argue that doing so is is possible, but not trivial.

First, attackers may try to use code-transformation techniques~\cite{moog2021}, such as obfuscation (which is a clear violation of the CWS policies~\cite{be-obf}), to conceal malicious functionalities. In these cases, our source-code features may be impacted; however, given that such features are analyzed by \jast{}, and given that \jast{} was designed to handle obfuscated code~\cite{JaSt}, it is safe to assume that attempting to evade our \texttt{source-code} and \texttt{combined classifiers} in this way is unlikely to succeed. Alternatively, attackers may try to embed malicious functionalities in segments of an extension that are not included in JavaScript (e.g., by manipulating the CSS~\cite{picazo2020after}). These attempts would be unnoticed by the \texttt{source-code classifier}; yet, such malicious extensions may still be detected by the \texttt{metadata} or \texttt{combined classifiers} because signals of maliciousness may be reflected in the features extracted from the extension's metadata. 

Attackers with knowledge of the features (and, particularly, the metadata-related) used by our classifiers may try to avoid detection by inducing manipulations that alter features important for a correct classification. For instance, being aware of our findings in Table~\ref{tab:keywords} or in Table~\ref{tab:feature_importance} may induce attackers to avoid mentioning terms such as ``WALLPAPER'' in an extension's summary, since this is a common occurrence across the malicious extensions in our datasets (as also found in~\cite{Sanchez2022}). Such attempts may make detection harder for the \texttt{metadata classifier}, but would not impact the source-code features used by the \texttt{source-code} and \texttt{combined classifier}. 

In other words, evading the \texttt{combined classifier} is not simple because, by analyzing both metadata and source-code features, it can cover the blind spots of the \texttt{metadata} and \texttt{source-code classifiers}. Moreover, it is possible to use fuzzying or feature-fusion methods to make such evasion attempts more difficult~\cite{ayantayo2023network,apruzzese2022mitigating}; or even preventing them altogether via feature-removal~\cite{smutz2012malicious}. Nevertheless, the worst threat against our detectors are ``adversarial ML attacks''~\cite{apruzzese2022mitigating}, which can be staged with full-knowledge of (and/or complete access to) the targeted classifier. In these cases, attackers may guess the manipulation that guarantees evasion---but in the feature space. However, due to the ``inverse-mapping'' problem, it is not trivial to understand how to manipulate the extension \textit{in the problem space} so as to induce a feature-representation that leads to evasion~\cite{cortellazzi2025intriguing}. 

Assessing the effectiveness and feasibility of all these different adversarial use-cases is a potential avenue for future work---which are facilitated by the full release of our resources~\cite{repository}.

\section{Related Work}
\label{sec:related}

\noindent
We discuss prior work on the security of browser extensions, and then highlight the major differences with competing papers, emphasizing our novelty.

\subsection{Browser Extensions and Security (\& Privacy)}
\label{subsec:extension_security}

\noindent
Early works in this domain are the 2012 papers by Carlini et al. and Liu et al., who assessed the effectiveness of Chrome security mechanisms by respectively semi-manually reviewing only 100 extensions~\cite{Carlini2012} and designing practical attacks through extensions~\cite{Liu2012}. 
After these seminal works, abundant prior literature showed that \textit{malicious extensions} can, e.g., track users~\cite{Weissbacher2017}, inject ads~\cite{kurt2015}, hijack Facebook accounts~\cite{Jagpal2015}, facilitate malvertising~\cite{Xing2015}, conceal botnets~\cite{perrotta2018botnet}, or exfiltrate sensitive user information~\cite{Aggarwal2018, Chen2018, Pantelaios2020}. These efforts confirm that malicious extensions pose a plethora of risks to (millions of) Web users.

Among the most recent related studies, Hsu et al.~\cite{Hsu2024} investigate the prevalence of ``security-noteworthy extensions'' in the CWS. However, Hsu et al.~\cite{Hsu2024} do not focus on detecting such security threats: aside from showing that malicious extensions have some similarities (which is an assumption we use to base our research), Hsu et al.~\cite{Hsu2024} do not propose any detector {(plus there is no ``open world'' assessment as we did in \Cref{sec:open}), and their findings are entirely based on ground truth provided by Google (via their proprietary and closed-source mechanisms).}

We focus on \textit{malicious} extensions, i.e., extensions designed by malicious actors to harm victims. However, there are other classes of extensions that are source of security or privacy concerns. In particular, \textit{fingerprintable} extensions (covered by, e.g.,~\cite{Agarwal2024, Karami2020, Starov2017, Starov2019, Solomos2022-2,Laperdrix2021,Sjosten2017,Solomos2022,SanchezRola2017,Sjosten2019,Trickel2019,Karami2022}), can be abused by attackers to (uniquely) identify users having a specific set of extensions installed on their browser, leading to, e.g., user tracking.
Also, \textit{vulnerable} extensions (covered by, e.g.,~\cite{Calzavara2015,Some2019,fass2021doublex,Yu2023,Kim2023,Buyukkayhan2016}), are designed by well-intentioned developers, but attackers can exploit their elevated privileges to put the security and privacy of the extension users at risk. These types of extensions (fingerprintable/vulnerable) are outside our scope (as also remarked in §\ref{subsec:scope}).

\subsection{Detecting Malicious Browser Extensions}
\label{subsec:detecting_extensions}

\noindent
Many works proposed ways to ``detect'' malicious extensions. After providing a summary, we carry out a systematic literature review to underscore our unique traits.

Closely related to us, Jagpal et al.\ developed WebEval, a semi-automatic system that is (or was) used by Google to analyze $\approx$99k extensions between 2012--2015~\cite{Jagpal2015}. 
However, WebEval is \textit{not publicly available}, and also \textit{requires regular inputs from human experts} to limit false negatives; furthermore, the findings of~\cite{Jagpal2015} are \textit{almost 10 years old}, and the landscape of browser extensions has substantially changed since 2015~\cite{Hsu2024}.
Equally close to us, Wang et al.~\cite{Wang2018} leveraged 51 static and dynamic features to train an SVM classifier to detect malicious Chrome extensions, and achieve 96\% accuracy on a dataset of nearly 5k extensions collected in 2015--2016. However, they handpicked their features also on the test set, hence overfitting; furthermore, they do not carry out a temporal evaluation---which we proved is necessary to ensure that the results reflect the real world. 
Also related, in 2018, Aggarwal et al.~\cite{Aggarwal2018} developed a classifier to detect ``spying'' extensions, and achieve 90\% precision on a dataset of 43k extensions (with manually-verified ground truth). However, their experimental pipeline also does not account for the temporal axis, leading to overestimation of their detector's performance in the long term. 

\begin{table}[h]
    \centering
    \resizebox{0.65\columnwidth}{!}{
        \begin{tabular}{l|c?c|c|c|c|r}
            \toprule
            \begin{tabular}{c} \textbf{Paper} \\ \textbf{(1st author)} \end{tabular} & 
            \textbf{Year} & 
            \begin{tabular}{c} \textbf{Considered} \\ \textbf{Browser} \end{tabular} & 
            \begin{tabular}{c} \textbf{Code} \\ \textbf{Public?} \end{tabular} & 
            \begin{tabular}{c} \textbf{Detect} \\ \textbf{Malicious?} \end{tabular} & 
            \begin{tabular}{c} \textbf{Supervised} \\ \textbf{ML used?} \end{tabular} & 
            \begin{tabular}{c} \textbf{Number of} \\ \textbf{Extensions} \end{tabular} \\
            \midrule
            Kapravelos~\cite{Kapravelos2014} & 2014 & C & & \cmark{} & \xmark{} & 48k\\
            Jagpal~\cite{Jagpal2015} & 2015 & C & & \cmark{} & \cmark & 100k\\
            Kurt~\cite{kurt2015} & 2015 & C,F,E & & \cmark{} & \xmark{} & 50k \\
            Xing~\cite{Xing2015} & 2015 & C & & \cmark{} & \xmark{} & 18k \\
            Buyukkayhan~\cite{Buyukkayhan2016} & 2016 & F & & \cmark{} & \xmark{} & 323 \\
            Sanchez-Rola~\cite{SanchezRola2017} & 2017 & C,F,S & & & & 21k\\
            Starov~\cite{Starov2017} & 2017 & C & & & & 10k\\ 
            Starov~\cite{Starov2017-2} & 2017 & C & & \cmark{} & \xmark{} & 10k\\
            Weissbacher~\cite{Weissbacher2017} & 2017 & C,F & & \cmark{} & \cmark & 43k \\
            Aggarwal~\cite{Aggarwal2018} & 2018 & C & & \cmark{} & \cmark & 178k\\
            Chen~\cite{Chen2018} & 2018 & C,O & \href{https://github.com/wspr-ncsu/mystique}{\cmark} & \cmark{} & \xmark{} & 19k \\
            Trickel~\cite{Trickel2019} & 2019 & C & \href{https://github.com/sefcom/cloakx}{\cmark} & & & 11k \\
            Sjosten~\cite{Sjosten2019} & 2019 & F,C & & & & 11k \\
            Some~\cite{Some2019} & 2019 & F,C,O & & & & 79k\\
            Starov~\cite{Starov2019} & 2019 & C & \href{https://github.com/plaperdr/extension-access-control}{\cmark} & & & 58k \\
            Karami~\cite{Karami2020} & 2020 & C & & & & 29k\\
            Pantelaious~\cite{Pantelaios2020} & 2020 & C & \href{https://github.com/wspr-ncsu/extensiondeltas}{\cmark} & \cmark{} & \xmark{} & 209k \\
            Borgolte~\cite{borgolte2020understanding} & 2020 & C,F & \href{https://github.com/noise-lab/privacy-extensions}{\cmark} & & & 8 \\
            Laperdrix~\cite{Laperdrix2021} & 2021 & C & \href{https://github.com/plaperdr/fingerprinting-in-style}{\cmark} & & & 116k \\
            Fass~\cite{fass2021doublex} & 2021 & C,F & \href{https://github.com/Aurore54F/DoubleX}{\cmark} & & & 154k \\
            Solomos~\cite{Solomos2022} & 2022 & C & \href{https://github.com/kostassolo/dangers-of-human-touch}{\cmark} & & & 40k \\
            Karami~\cite{Karami2022} & 2022 & C & \href{https://github.com/SimulacrumExtension/Simulacrum}{\cmark} & & & 5.7k \\

            Solomos~\cite{Solomos2022-2} & 2022 & C & & & & 38k \\
            Picazo-Sanchez~\cite{Sanchez2022} & 2022 & C & & \cmark{} & \cmark & 159k \\
            
            Agarwal~\cite{Agarwal2022} & 2022 & C,F & \href{https://github.com/shubh401/black_canary}{\cmark} & & & 186k \\ 
            Kim~\cite{Kim2023} & 2023 & C,F,S & \href{https://github.com/compsec-snu/exthand}{\xmark} & & & $<$100\\
            Bui~\cite{buidetection} & 2023 & C & & & & 47k \\
            Moreno~\cite{moreno2023chrowned} & 2023 & C & \href{https://github.com/josemmo/chrowned}{\cmark} &  & & --- \\
            Yu~\cite{Yu2023} & 2023 & C & \href{https://github.com/CoCoAbstractInterpretation/CoCo}{\cmark} & & & 145k \\
            Xie~\cite{Xie2024} & 2024 & C & \href{https://github.com/BEESLab/Arcanum/}{\cmark} & & & 113k \\ 
            Nayak~\cite{nayak2024experimental} & 2024 & C & & \cmark{} & \xmark{} & 160k \\
            Olsson~\cite{Olsson2024} & 2024 & C & & & & 115k \\
            Hsu~\cite{Hsu2024} & 2024 & C & & & & 226k \\
            Solomos~\cite{solomos2024harnessing} & 2024 & C & & & & 60k \\
            Agarwal~\cite{Agarwal2024} & 2024 & C,F & \href{https://github.com/raider-ext/raider}{\cmark} & & & 88k\\
            \midrule
            This work & 2024 & C & \cite{repository} & \cmark{} & \cmark & 106k \\
            \bottomrule
        \end{tabular}
    }
    \caption{\textbf{State of the Art.} Papers having ``extension'' in the title published in 2014--2024 in top-tier conferences. For each paper, we report the considered browser (F=Firefox, C=Chrome, O=Opera, S=Safari, E=Explorer), if the code is public (if so, the \cmark{} links to the repository; the link for~\cite{Kim2023} is broken), if it focuses on the ``detection'' of malicious extensions (and, if so, if supervised ML is used), and the number of extensions analyzed.}
    \label{tab:related}
\end{table}

Some works also focused on detecting malicious extensions via heuristics or dynamic analyses supported by \textit{unsupervised} ML---which is a different category of ML-based techniques that, despite not requiring labeled data, necessitate custom-defined rules/heuristics to detect malicious extensions.
Such works typically find only a small number of malicious extensions via manual verification. For instance, Kapravelos et al.~\cite{Kapravelos2014} analyzed 48k Chrome extensions and found 130 malicious ones; Wang et al.~\cite{Wang2012} analyzed 2.5k Firefox extensions, finding no malicious behavior. 
Olsson et al.~\cite{Olsson2024} used fake reviews on the CWS to identify 86 malicious extensions among 115k. 
Notably, Pantelaios et al.~\cite{Pantelaios2020} leveraged the evolution of extensions' ratings to {identify potentially malicious clusters, finding 143 malicious extensions out of over 206k.\footnote{In~\cite{Pantelaios2020}, it is stated that ``our system requires an amount of manual analysis both for identifying the true positives of malicious extensions that can serve as seeds for identifying others, but also to differentiate between true positives and false positives after the clustering-deltas step. We therefore see our system as a helping tool for security analysts''---thereby admitting that their tool has limitations which we sought to overcome through the application of supervised-ML methods.} Nonetheless, some of our metadata features are inspired by~\cite{Olsson2024,Pantelaios2020}, albeit we use these in a supervised-ML context.

\subsection{Systematic Literature Analysis}
\label{subsec:systematic}
\noindent
We carry out a systematic literature review to highlight some relevant differences w.r.t. prior work, but also to corroborate our previous claims (see Sections~\ref{sec:introduction} and~\ref{subsec:scope}). Specifically, we show the \textit{lack of public source code}, preventing direct comparison; and the \textit{non-reliance on supervised-ML methods}, preventing assessments of ``concept drift'' (which is a problem pertaining to supervised ML~\cite{gama2014survey}).

\pseudoparagraph{Method} We consider \textit{all papers published in top-tier conferences having ``extensions'' in the title}. We considered the 2014--2024 editions of: USENIX Sec, TheWebConf, IEEE S\&P / EuroS\&P, ACM CCS / AsiaCCS, and NDSS. We found 35 papers that matched our criteria, which were then inspected independently by two researchers. Specifically, we first analyzed the abstract to determine if the paper was about ``detection of malicious browser extensions'' (we resolved cases of disagreement via discussions). Then, we did whole-text keyword searches for three common ML-related terms (``machine learning'', ``training'', ``deep learning''), and if we found more than 3 matches we inspected the paper to see if some form of supervised ML\,/\,classifier was used in the detection procedure. For the code availability, we looked for links to repositories, and checked if they were still active.

\pseudoparagraph{Results} The results are in Table~\ref{tab:related}, in which we also report the considered browser and size of the dataset (in terms of unique extensions) used in the assessment (if any). At a high level, we see that, with the exception of one paper~\cite{Buyukkayhan2016}, all other works consider Chrome, confirming our choice to focus on this Web browser; we also see that, in terms of dataset size, only eight works (i.e.,~\cite{Hsu2024,Agarwal2022,Sanchez2022,fass2021doublex,Pantelaios2020,Aggarwal2018, Yu2023,nayak2024experimental}) carry out a substantially larger (i.e., $>$30\%) assessment than ours. Let us focus on the most crucial factors of our systematic literature review:
\begin{itemize}[leftmargin=*]
    \item \textit{(Not) Open Source:} only 15 (out of 35) papers publicly release their code; however, the repository of one of these (i.e.,~\cite{Kim2023}) is not active (as of July 2025). This result echoes that of a recent work revealing that less than half of papers published at top-tier conferences release their artifacts~\cite{olszewski2023get}.
    \item \textit{Different Focus:} only 12 (out of 35) papers focus on the ``detection of malicious extensions''; other works focus on different classes of security-related problems of the browser-extension ecosystem (e.g.,~\cite{Xie2024} focuses on detecting vulnerable extensions, whereas~\cite{moreno2023chrowned} shows a vulnerability of Chromium-based browsers, and do not carry out any analysis of browser extensions).
    \item \textit{Limited Consideration of Supervised ML:} of the 12 papers that focus on the detection of malicious browser extensions, only 4 (i.e.,~\cite{Jagpal2015,Weissbacher2017,Aggarwal2018,Sanchez2022}) attempt to do so by relying on some form of supervised-ML technique (e.g.,~\cite{Pantelaios2020} uses unsupervised methods)--and their code is not public. 
\end{itemize}
We also looked for occurrences of ``concept drift'': no paper among these 35 mentioned this term.

\pseudoparagraph{Consequences} Let us explain how the aforementioned results affect our research. First, \textit{there is no work that uses supervised ML to detect malicious browser extensions and which publicly shares its implementation} (e.g.,~\cite{Sanchez2022} uses supervised ML, but there is no code available). This result implicitly prevents any sort of fair comparison with a previously proposed method; and it also reinforces our choice to consider \jast{}~\cite{JaSt} as the most valid baseline (despite not being designed to work on browser extensions). Second, we observe that there are two works (i.e.,~\cite{Pantelaios2020, Chen2018}) for which an artifact is available and that focus on the detection of malicious browser extensions, but not by means of supervised-ML. However, we cannot compare against~\cite{Pantelaios2020} due to a substantially different setup: our datasets include a single version for each extension, whereas the method proposed in~\cite{Pantelaios2020} require {\small \textit{(i)}}~multiple versions of each extensions, and---crucially---that {\small \textit{(ii)}}~only the last version of a malicious extension is malicious, whereas all the previous ones are benign. Due to lack of ground truth on benign extensions, we cannot be sure if the previous version of the malicious extensions contained in our datasets are truly benign (indeed, Hsu et al.~\cite{Hsu2024} found that malicious extensions can stay in the CWS for years). Nevertheless, we tried using the open-source implementation of the method proposed in~\cite{Chen2018}, but we did not succeed. This negative result echoes a remark made in~\cite{Xie2024}, stating that \textit{some prior tools are ``incapable of operating in a modern context'' due to advances in ``modern browsers, the expressiveness of extensions APIs, and the web itself''.} We also contacted the authors of~\cite{Kapravelos2014,Chen2018}, who confirmed that these tools are not possible to compare against anymore due the significant evolution of the Chrome browser and its extension ecosystem, which led to deprecated APIs not supported by current versions of Chrome.
We believe that this is yet another finding supporting our claim that the browser-extension ecosystem is affected by concept drift.

\section{Conclusions and Lessons Learned}
\label{sec:conclusions}

\noindent
We performed a security assessment, of the extensions published on the Chrome Web Store (CWS).
We collected 7,140 known malicious extensions {(removed from the CWS by Google) and} provided by Chrome-Stats, along with 63,598 benign extensions available on the CWS and last updated before 2023.
We devised and implemented three supervised-ML-based classifiers (including one based entirely on prior work~\cite{JaSt}) using features we extracted from extensions' metadata and source code. These classifiers \textit{perform well} {in a ``lab setting''}, with an accuracy of 98\% and can analyze an extension in less than 1s end-to-end (even on commodity hardware).
Subsequently, we show that these detectors are \textit{underwhelming in an open-world setting} {by using them to analyze an extra 35k extensions from the CWS with unverified ground truth}.
Despite enabling us to identify 68 malicious extensions (which were still present in the CWS at the time of our analysis, and which cumulatively affected over 13M users), our detectors flag over 1k extensions as malicious---which is an unrealistically high number that may conceal many false positives.
We also provide evidence that \textit{commercial detectors perform poorly} to detect known malicious extensions, i.e., \textit{existing methods may give a false sense of security}---confirmed by our tests showing that, e.g., the detectors of VirusTotal have a false-negative rate over 97\%.
Finally, we show that the browser extension ecosystem is \textit{affected by concept drift}.
Altogether, our results serve as a foundation for researchers and practitioners to improve the detection of malicious extensions in the CWS, which \textit{affect millions of end-users}.

We can hence derive \textbf{lessons learned} addressed to the entire browser-extension community:
\begin{itemize}[leftmargin=*]
    \item First, \textit{ML evaluations can be misleading}: a detector performing well in a research setting is no guarantee of practical applicability in an open-world scenario. In fact, to the best of our knowledge, we are the first to demonstrate that the browser extension ecosystem is affected by concept drift. Hence, from the perspective of a researcher, it is crucial to train and test models via time-aware evaluations for real-world assessments. From a practical perspective, we underscore that detectors' performance deteriorate quickly and need constant updates with new labeled data, i.e., both benign and malicious. This, however, leads to another problem: labeling extensions is hard. As we showed, existing tools, e.g., VirusTotal, are unreliable.
    \item Second, to improve the security of the browser extension ecosystem, \textit{multiple stakeholders must cooperate}. End users should be willing to report suspicious extensions. Security researchers need to design more reliable tools and carry out more realistic security assessments. 
    Maintainers of extension galleries should be more active in fighting these threats: e.g., the CWS contains numerous malicious extensions made by developers known to publish dangerous extensions. 
    \item Third, to spearhead development of appropriate solutions, we have (for the first time) carried out experiments on \textit{active learning} (described in Section~\ref{subsec:active}): our findings (in Figure~\ref{fig:activelearning}) suggest that it is possible to optimally improve the performance of our classifiers by retraining them on a monthly basis on a small number (we found 15 is a sweet spot) of correctly labeled extensions.
\end{itemize}
Finally, our findings mostly pertain to Chrome. However, our approach {(and, likely, our conclusions)} is also applicable to other browsers, such as Firefox and Microsoft Edge, with minor adjustments, e.g., by leveraging data from Firefox-Stats~\cite{FirefoxStats} and Edge-Stats~\cite{EdgeStats}.

\begin{acks}
We would like to thank ChromeStats~\cite{ChromeStats}, and more specifically Hao Nguyen, for providing us with Chrome extensions, their metadata, and additional information or support upon request. We also thank the anonymous TWEB reviewers for their constructive reviews and helpful feedback.
Parts of this research was funded by the Hilti foundation.
\end{acks}

\bibliographystyle{ACM-Reference-Format}
%\bibliography{bibliography}

\appendix

\section{Extra Figures and Tables}
\label{app:figtables}

\noindent
We provide additional figures and tables to support our claims.

\begin{figure}[!htbp]
    
    \centering
    \includegraphics[width=.75\columnwidth]{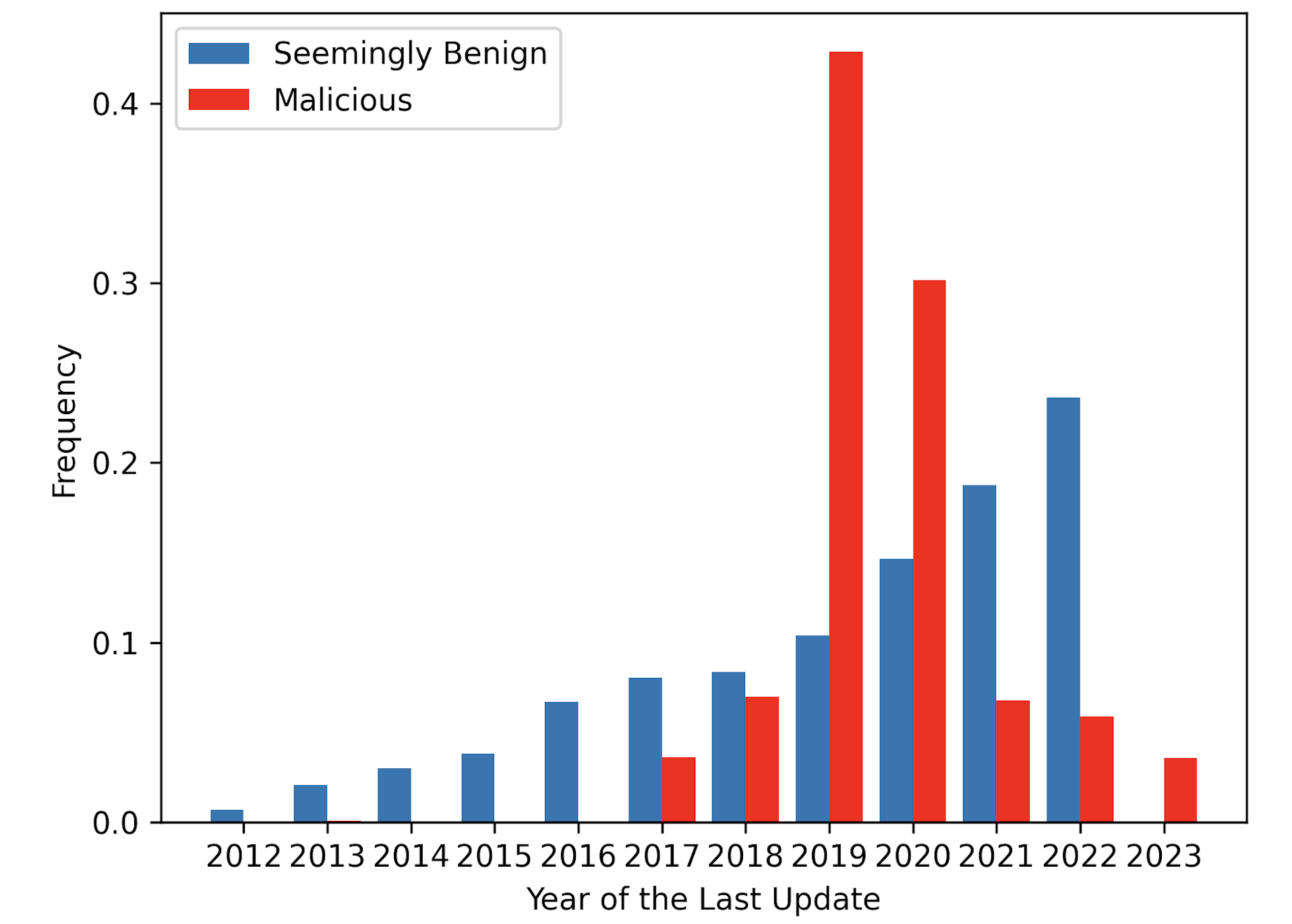}
    \vspace{-2mm}
    \caption{Temporal distribution of the extensions in \smallseta. (N.b.: all extensions in \smallsetb\ have had their last update in 2023)} 
    \label{fig:distribution}
    \vspace{-3mm}
\end{figure}

\begin{figure}[!htbp]
    \centering
    \includegraphics[width=.95\columnwidth]{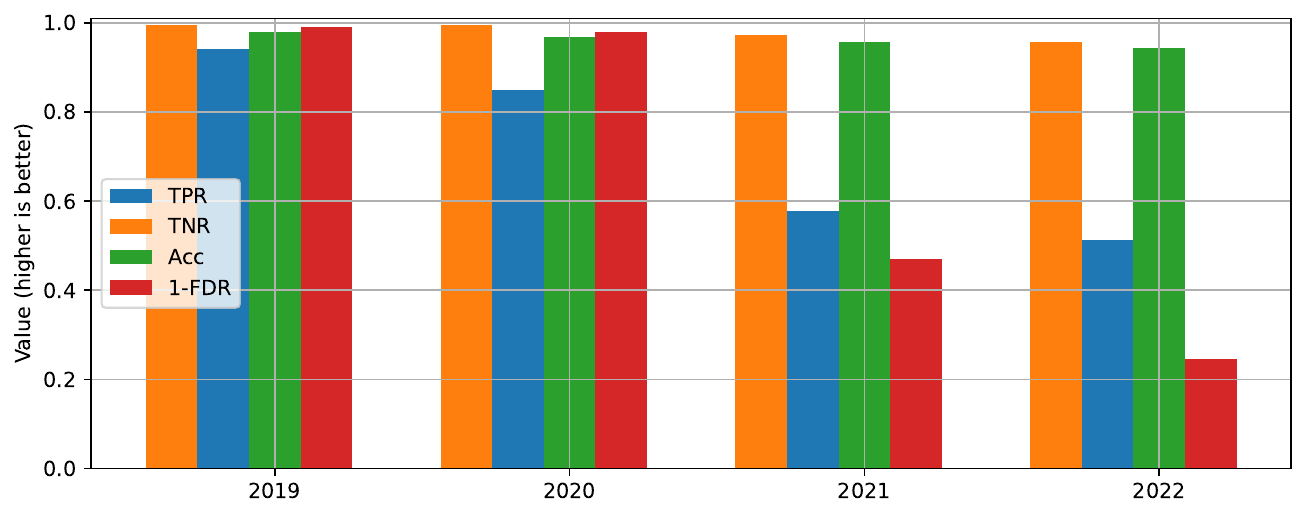}
    \vspace{-2mm}
    \caption{Performance over time. We test the \smallmeta{} on the extensions updated every year from 2019--2022 after training it on the extensions published or updated in the previous years.}
    \label{fig:conceptdrift-meta}
\end{figure}

\end{document}